\documentclass[acmsmall]{acmart}

\AtBeginDocument{%
  }

\setcopyright{cc}
\setcctype{by}
\acmDOI{10.1145/3830439.3831272}
\acmYear{2026}
\copyrightyear{2026}
\acmISBN{979-8-4007-2869-3/2026/08}
\acmConference[Haskell '26]{Proceedings of the 19th ACM SIGPLAN International Haskell Symposium}{August 24--29, 2026}{Indianapolis, IN, USA}
\acmBooktitle{Proceedings of the 19th ACM SIGPLAN International Haskell Symposium (Haskell '26), August 24--29, 2026, Indianapolis, IN, USA}
\acmSubmissionID{icfpws26haskellmain-p8-p}
\received{2026-05-22}
\received[accepted]{2026-06-30}

\usepackage{cleveref}
\usepackage{xcolor}
\usepackage{subcaption}
\usepackage{siunitx}
\newcommand{\ch}[1]{Cloud Haskell}
\newcommand{\mch}[1]{CloudMicroHaskell}
\newcommand{\paper}[1]{paper} 
\newcommand{\inlinehaskell}[1]{\lstinline[style=mchhaskell,keepspaces=true]|#1|}
\newcommand{\inlinecombformat}[1]{\lstinline[style=combformatbase,keepspaces=true]|#1|}
\newcommand{\ms}[1]{MicroSync}

\newcommand{\lennart}[1]{\textcolor{green}{[Lennart: #1}]}

\usepackage{listings}
\lstdefinestyle{haskell}{
  frame=none,
  xleftmargin=2pt,
  stepnumber=1,
  numbers=left,
  numbersep=5pt,
  numberstyle=\ttfamily\tiny\color[gray]{0.3},
  belowcaptionskip=\bigskipamount,
  alsoletter={=<>-|{}},
  captionpos=b,
  escapeinside={*'}{'*},
  keywordstyle=[2]{\color{teal}},
  keywordstyle=[3]{\color{violet}},
  keywordstyle=[1]{\bf\color{purple}},
  keywordstyle=[4]{\color{blue}},
  keywords=[3]{INLINE, PRAGMA, NOINLINE, RULES, SPECIALISE, forall}, 
  keywords=[4]{Int, Bool, Double, Maybe, Either, Monad, Map, String, ByteString, IO, Proxy, KnownSymbol, Show, Read, IORef, Num, True, False, MVar, Word32, Word64, Property},
  morekeywords=[1]{do, if, then, else, case, of, class, data, newtype, instance, where, deriving, import, let, in, module, qualified, type},
  otherkeywords={<-, ->, <>, ::, \$, [, ],/,\\, \,, =, =>, ==>, |, \{, \}, (, ), -\#, \#-, @},
  morekeywords=[2]{-\#, \#-, \{, \#, \}},
  morekeywords=[4]{@},
  tabsize=2,
  comment=[l]{--},
  commentstyle=\color{teal}, 
  stringstyle=\ttfamily,
  showspaces=false,
  delim=[s][\ttfamily\color{orange}]{"}{"},
  columns=flexible,
  basicstyle=\small\ttfamily,
  showstringspaces=false,
}

\lstdefinestyle{erlang}{
  frame=none,
  xleftmargin=2pt,
  stepnumber=1,
  numbers=left,
  numbersep=5pt,
  numberstyle=\ttfamily\tiny\color[gray]{0.3},
  belowcaptionskip=\bigskipamount,
  alsoletter={?@-},
  captionpos=b,
  escapeinside={*'}{'*},
  keywordstyle=[2]{\color{teal}},
  keywordstyle=[3]{\color{violet}},
  keywordstyle=[1]{\bf\color{purple}},
  keywordstyle=[4]{\color{blue}},
  keywords=[1]{after,begin,catch,case,end,fun,if,let,of,receive,try,when},
  keywords=[3]{-module,-export,-import,-define,-record,-spec,-type,-callback,-opaque},
  keywords=[4]{atom,boolean,float,integer,list,tuple,map,pid,port,reference},
  otherkeywords={->,<-,=>,=,!,:,::,;,.,\,,++,--,orelse,andalso,'},
  tabsize=2,
  comment=[l]{\%},
  commentstyle=\color{teal},
  stringstyle=\mdseries\rmfamily,
  showspaces=false,
  delim=[s][\ttfamily\color{orange}]{"}{"},
  columns=flexible,
  basicstyle=\small\sffamily,
  showstringspaces=false,
}

\lstdefinestyle{mchhaskell}{
  style=haskell,
  xleftmargin=2.4em,
  numbersep=0.6em,
  morekeywords=[4]{
    Flags,NodeConfig,NodeId,Node,Pid,ProcessM,MatchM,MonitorAction,
    ProcessDied,MonitoredProcessDied,ExitReason,Identifiable,
    Typeable,NFData,MonadIO,
    TrapExit,Succumb,
    ExitNormal,ExitShutdown,ExitKill,ExitOther,
    Just,Nothing, Server, ChildSpec, SupervisorSpec, Restart, Strategy,
    Permanent, Transient, Temporary, OneForOne, OneForAll, Inc, Get, Res,
    ServerSpec, CastRequest, CallRequest, NodeState, SockAddr, Socket, Mailbox, ThreadId,
    PendingReceive, NodeCommand, Local, Command, NonLocal, PeerCommand, Connect, Deliver,
    PeerConnect, PeerDeliver, Mail, WireCommand, WireConnect, WireDeliver, Register,
    SpawnOn, WireSpawn, PeerSpawn, WireSpawnAck, PeerSpawnAck, HeapMail, Static, Closure,
    MkClosure, Serializable, PingPong, Ping, Pong, Binary, Apply, Reset, TypeDescribable,
    TypeRep, Ptr, Word8, EncodedMail, WireMonitor, WireProcessDied, ProcessExit, WireExit,
    Div, Shutdown, Queue, CommandQueue
  },
  keywordstyle=[5]{\color{purple!70!black}},
  morekeywords=[5]{
    mkNodeConfig,runNode,runProcessM,
    connect,pconnect,node,nodes,pnode,pnodes,self,
    spawn,call,send,expect,monitor,terminate,exit,
    match,matchIf,matchUnknown,receiveWait,receive,receiveTimeout,
    register,unregister,whois
  }
}

\lstdefinestyle{combformatbase}{
  basicstyle=\ttfamily,
  columns=fullflexible,
  keepspaces=true,
  showstringspaces=false,
  sensitive=true,
  alsoletter={'.},
  otherkeywords={==,+,*,-,@,\#,_,:},
  keywordstyle=[1]{\bfseries\color{purple!70!black}},
  morekeywords=[1]{S,K,I,B,C,C'},
  keywordstyle=[2]{\bfseries\color{blue!70!black}},
  morekeywords=[2]{@},
  keywordstyle=[3]{\color{orange!85!black}},
  morekeywords=[3]{\#,_,:},
  keywordstyle=[4]{\color{teal!70!black}},
  morekeywords=[4]{v8.4},
  keywordstyle=[5]{\color{violet!80!black}},
  morekeywords=[5]{==,+,*,-},
  xleftmargin=8pt,
  xrightmargin=8pt,
}

\lstdefinestyle{combformat}{
  style=combformatbase,
  frame=single,
  framerule=0.3pt,
  rulecolor=\color{black!20},
  backgroundcolor=\color{black!2},
  xleftmargin=2pt,
  framexleftmargin=4pt,
  framexrightmargin=4pt,
  numbers=none,
  basicstyle=\small\ttfamily,
  breaklines=true,
}

\begin{document}

\title{CloudMicroHaskell}
\subtitle{Direct-Style Distributed Haskell via Runtime Graph Serialisation}

\author{Robert Krook}
\orcid{0000-0003-3619-2975}
\affiliation{%
  \institution{Chalmers University of Technology and University of Gothenburg}
  \department{Department of Computer Science and Engineering}
  \city{Gothenburg}
  \country{Sweden}
}
\email{krookr@chalmers.se}

\author{Lennart Augustsson}
\orcid{0009-0008-6894-4020}
\affiliation{%
  \institution{Unaffiliated}
  \city{Gothenburg}
  \country{Sweden}
}
\email{lennart@augustsson.net}

\begin{abstract}
\ch{} brings Erlang-style distributed programming to Haskell, but its treatment of mobile code exposes a difficult boundary in the source-level API.
Remote processes must be expressed as static closures, messages must satisfy serialization constraints, and participating nodes are assumed to share the relevant code.
  
This paper explores a different design point.
We present \mch{}, a \ch{}-style library built on MicroHaskell, whose runtime represents both code and data as a combinator graph.
When a process or message crosses a node boundary, \mch{} serializes the reachable graph directly.
As a result, remote spawning can be written in direct fashion: process bodies may capture variables from their surrounding scope, and messages may contain ordinary values, including functions, without programmer-written closure conversion.
  
We describe the implementation of the \mch{} node runtime, including remote spawn, message delivery, monitors, exit propagation, and library implementations of generic servers and supervisors.
We evaluate the system with process/message benchmarks, a distributed work-pool benchmark, a file-synchronization case study, and a heterogeneous deployment on microcontrollers.
The results show that runtime graph serialization makes the \ch{} programming model substantially more direct, while also making the tradeoff explicit: some guarantees enforced by \ch{}'s source-level types become dynamic checks, and programmers must be aware of laziness and runtime-owned resources when moving graphs between nodes.
\end{abstract}



\maketitle

\section{Introduction}

\ch{}~\cite{DBLP:conf/haskell/EpsteinBJ11} is a Haskell framework for writing distributed, concurrent programs in the style of Erlang.
Erlang's programming model emphasizes concurrency, with lightweight processes that share no memory. Instead, inter-process communication is done via message passing.
This programming model greatly aids in writing distributed programs that execute on and communicate over a network.

The primary benefit of \ch{} over Erlang is Haskell's strong type system.
With it, the API of \ch{} can be described in a very clear and expressive way.
It is short and (mostly) elegant. Furthermore, \ch{} extends Erlang's untyped message-passing machinery with typed channels, so that communicating processes can exchange messages whose format is known at compile time.
Erlang, being dynamically typed, cannot statically enforce a message-passing protocol between processes.

These advantages come from making distribution explicit in the API, and that explicitness has practical consequences.
The implementation assumes that nodes in the network run a very uniform software environment, and the machinery required to handle closures is necessarily visible to the programmer.
We illustrate this with the direct-style program one might like to write as:

\begin{lstlisting}[style=mchhaskell]
  sendFunc :: Pid -> Int -> ProcessM ()
  sendFunc p x = send p (\y -> x + y + 1)
\end{lstlisting}

The above example sends a lambda to a recipient process, which captures a free variable (\inlinehaskell{x}).
\ch{} uses the type system to enforce that values are serialized before transmission, but note that the type of the function, \inlinehaskell{Int -> Int}, says nothing about the type of the free variable.
It is therefore not possible to provide an ordinary serializer for functions from the function type alone.

To get around this, a mechanism for \inlinehaskell{Static} values was devised.
Explained briefly, a value is considered static if it is defined at the top level and refers to only bound variables or other variables defined at the top level.
Furthermore, when a static function is transmitted to a remote node, it must be explicitly closure converted, whereby the environment of the function is populated with its (serialized) free variable.
The function is responsible for deserializing its environment itself.
To implement the above example, the programmer must write:

\begin{lstlisting}[style=mchhaskell,firstnumber=1,aboveskip=0pt,belowskip=0pt]
  sendFunc :: Pid -> Int -> ProcessM ()
  sendFunc p x = send p clo
    where
      -- closures can be serialized
      clo :: Closure (Int -> Int)
      clo = MkClosure (static sfun) (encode x)

  sfun :: ByteString -> Int -> Int
  sfun bs y = let x = decode bs
              in  x + y + 1
\end{lstlisting}

The previously intuitive and brief program has now been rewritten in a style with a significantly higher cognitive cost.
Additionally, a static value is assumed to exist on both machines (the one running the sending process and the one running the receiving process), and a closure only contains a static reference to it (together with its environment).
While it seems like a function is being sent, what is actually being sent is a static reference to it and its environment.

While this is a clever solution to a tricky problem, we note that there are strict requirements placed on participating machines in the network, namely that they must run binaries that contain the same static symbols.

In this \paper{}, we ask what \ch{} might look like if the serialization boundary were moved out of the source-level API and into the runtime representation of the program.
We present \mch{}, an exploration of the \ch{} programming model using the MicroHaskell compiler~\cite{augustsson2024microhs} (MicroHs for short).
MicroHs is a Haskell compiler with a much smaller and more modest runtime system than GHC, the de facto Haskell compiler.
MicroHs emits a \textit{combinator graph} as its object code — a self-modifying byte code that represents both a program's data and code.
While the cost of this decision is a significant reduction in performance, there are several benefits which we hope to illustrate with this \paper{}.

Using MicroHs changes the tradeoffs in several useful ways, namely:
\begin{itemize}
  \item We use a simple compiler primitive to serialize values of (almost) any kind, including functions. The first variant of \inlinehaskell{sendFunc} in the introduction is directly runnable in \mch{}.
  \item Our implementation does not require every node in the network to agree beforehand on which static values should exist. Instead, nodes need only agree on a compatible \mch{} runtime and the combinator format understood by that runtime.
  \item \mch{} can serialize and transmit a message of (almost) any type, without requiring a serializable constraint. This reduces the boilerplate required to prepare a data type for transmission over a network.
\end{itemize}

This does not remove the distribution boundary, but relocates it.
The price is that this boundary is no longer enforced by Haskell's source-level types.
Deserialization is necessarily a dynamic claim that the received graph has the type expected by the receiver.
If that claim is wrong, the program may fail at runtime.
Because of that, \mch{} trades some of \ch{}'s static safety for a much simpler programming model.

The contribution is therefore not the actor API itself, which is deliberately conventional, but the consequences of making runtime graph serialization the mobility mechanism for that API.

\section{Programming in \mch{}}

In this section we present \mch{}.
At the surface, \mch{} looks much like \ch{}.
Programs create processes, identify remote nodes, and communicate by message passing.
This is deliberate.
The point of \mch{} is not to propose a new process calculus, but to explore what happens when the serialization boundary in \ch{} is moved out of the source-level API and into the MicroHs runtime.

The most important differences therefore appear in the two operations that move computation and data between nodes, \inlinehaskell{spawn} and \inlinehaskell{send}.
In \ch{}, these operations expose distribution through \inlinehaskell{Closure} and \inlinehaskell{Serializable} constraints.
In \mch{}, these operations accept ordinary process computations and ordinary values, relying instead on MicroHs's graph serializer (the serializer and deserializer are described in \cref{sec:implementation}).

We therefore focus on these two operations in the main text.
The surrounding API follows the familiar \ch{} and Erlang shape\footnote{The full API is described in \cref{appendix:API}.}.

The core difference is visible in the types:

\begin{lstlisting}[style=mchhaskell]
  spawn :: NodeId -> ProcessM () -> ProcessM Pid
  send  :: Identifiable process => process -> a -> ProcessM ()
\end{lstlisting}

In \ch{}, remote spawning requires a \inlinehaskell{Closure} and message passing requires a\\
\inlinehaskell{Serializable} constraint.
In \mch{}, the programmer supplies the computation or value directly.
If the target process or message crosses a node boundary, the runtime will serialize the reachable MicroHs graph, capturing both the code and environment.
This makes direct-style examples possible to execute, but it also means that some errors that \ch{} eliminates via the type checker become runtime failures in \mch{}.

The serialization primitive is intentionally very broad, but not universal.
Some things are owned by a particular runtime instance, such as \inlinehaskell{MVar}s and open file handles.
These cannot be serialized.
The exceptions are the standard handles \textit{stdin}, \textit{stdout}, \textit{stderr}, and the \textit{null pointer}.
These handles are recognized by the MicroHs runtime system.
They can be serialized and bound to the receiver's version during deserialization.
Similarly, the null pointer is always a zero-valued pointer, and can be serialized and deserialized as such.

The example below illustrates how \inlinehaskell{spawn} is used to spawn a concurrent process that captures free variables from the outer process:

\begin{lstlisting}[style=mchhaskell]
  data PingPong = Ping | Pong

  f :: ProcessM ()
  f = do
    s <- self
    n <- node
    n' <- (head . filter (/= n)) <$> nodes

    child <- spawn n' $ do
      Ping <- expect
      liftIO $ putStrLn "child received ping!"
      send s Pong

    send child Ping
    Pong <- expect
    liftIO $ putStrLn "parent received pong!"
\end{lstlisting}

The parent process first fetches its own process identifier by calling \inlinehaskell{self}.
It then spawns a child process on a remote node.
The child refers to \inlinehaskell{s}, which is bound in the parent.
When \inlinehaskell{spawn} crosses the node boundary, the child computation is serialized together with this reachable environment.
Additionally, we can send the \inlinehaskell{PingPong} data type without saying how to serialize it.

To make it even clearer that code is serialized and sent to a remote node, we show yet another example:

\begin{lstlisting}[style=mchhaskell]
  f :: ProcessM ()
  f = do
    s <- self
    n <- node
    n' <- (head . filter (/= n)) <$> nodes

    let interesting x = x `mod` 7 == 0 && x `mod` 11 == 0

    _ <- spawn n' $ do
      let r = length (filter interesting [1..1000000])
      r `seq` send s r

    r <- expect :: ProcessM Int
    liftIO $ print r
\end{lstlisting}

We define a local function \inlinehaskell{interesting :: Int -> Bool}, and reference it from the spawned process.
\inlinehaskell{interesting} will be included in the serialized graph that is sent to the remote node, and reconstructed on that node's heap.
The call to \inlinehaskell{seq} is intentional.
Since \mch{} serializes graphs, not values produced by a \inlinehaskell{Binary} encoder, sending \inlinehaskell{r} without first evaluating it could send a thunk back to the original node.
The result would still be correct, but the work would happen in the wrong place.

It is not just \inlinehaskell{spawn} that can be used to send functions to remote processes.
\inlinehaskell{send} uses the same underlying serializer, and will happily send functions.
Consider the example below:

\begin{lstlisting}[style=mchhaskell]
  data Apply = Apply (Int -> Int) Int Pid

  worker :: ProcessM ()
  worker = do
    Apply f x replyTo <- expect
    let y = f x
    y `seq` send replyTo y

  example :: ProcessM Int
  example = do
    me <- self
    n  <- node
    n' <- (head . filter (/= n)) <$> nodes

    workerPid <- spawn n' worker

    let offset = 1
    send workerPid (Apply (\x -> x + offset) 41 me)

    expect
\end{lstlisting}

The message sent to \inlinehaskell{workerPid} contains a function, and that function captures \inlinehaskell{offset}.
In \ch{}, this is exactly the case that requires static closures and explicit environment serialization.
In \mch{}, the function is just part of the message graph.
The receiver expects an \inlinehaskell{Apply} message, extracts the function, and applies it.

The rest of the library follows the familiar process-programming structure.
Nodes can be connected and queried, processes have identifiers and mailboxes, selective receive inspects messages by runtime type, monitors and exits make process failure observable, and a node-local registry lets long-running processes be found by name.
These operations are not the novelty of \mch{}, but they are necessary to make the core \inlinehaskell{spawn} and \inlinehaskell{send} design usable in larger programs.
Their full API is listed in \cref{appendix:API}.

\subsection{Process Behaviors as Libraries}

The examples above use \inlinehaskell{spawn} and \inlinehaskell{send} directly.
To test whether this style scales beyond our toy examples, we implement two familiar Erlang/OTP behaviors as ordinary \mch{} libraries, generic servers and supervisors.
These behaviors are not built into the runtime, but are implemented using the same process creation, message passing, monitors, and exits available to the programmer.

A generic server captures a common pattern whereby a process responds to requests and then enters a state where it is ready to respond to a subsequent request.
The process may also maintain a state which may be modified by a request, and the modified state must be used in subsequent requests.
The application-specific part is described by a record of callbacks:

\begin{lstlisting}[style=mchhaskell]
  data ServerSpec st callReq castReq reply = ServerSpec
    { setup      :: ProcessM st
    , handleCall :: st -> callReq -> ProcessM (st, reply)
    , handleCast :: st -> castReq -> ProcessM st
    , tearDown   :: st -> ProcessM ()
    }

  startServer :: NodeId -> ServerSpec st callReq castReq reply
              -> ProcessM (Server callReq castReq reply)
  -- run the server in the *current* process
  runServer :: ServerSpec st callReq castReq reply -> ProcessM ()

  call :: Server callReq castReq reply -> callReq -> ProcessM reply
  cast :: Server callReq castReq reply -> castReq -> ProcessM ()
\end{lstlisting}

\inlinehaskell{startServer} can start the server on a remote node, even though the server specification contains ordinary Haskell functions.
Internally, a server loop is spawned together with the supplied callbacks.
The programmer only implements the state transition logic, and the library handles the receive loop, reply routing, and state threading.

For example, a counter server can be described by giving its initial state and two handlers:

\begin{lstlisting}[style=mchhaskell]
  data CallRequest = Get
  data CastRequest = Inc | Reset
  
  counterSpec :: ServerSpec Int CallRequest CastRequest Int
  counterSpec = ServerSpec
    { setup      = return 0
    , handleCall = \st Get -> return (st, st)
    , handleCast = \st req ->
        case req of
          Inc   -> return (st + 1)
          Reset -> return 0
    , tearDown   = \_ -> return ()
    }
\end{lstlisting}

A client can then start the server remotely and interact with it through
a typed handle:

\begin{lstlisting}[style=mchhaskell]
  do counter <- startServer n counterSpec
     cast counter Inc
     cast counter Inc
     i <- call counter Get
     liftIO $ print i
\end{lstlisting}

Supervisors capture a second common pattern, where a supervisor process keeps a set of child processes alive according to a restart policy.
Again, the implementation is library code.
A supervisor starts its children using \inlinehaskell{spawn}, monitors them, and reacts to \inlinehaskell{ProcessDied} messages by consulting its policy.

\begin{lstlisting}[style=mchhaskell]
data ChildSpec = ChildSpec
  { childName    :: String
  , childStart   :: ProcessM ()
  , childNode    :: Maybe NodeId
  , childRestart :: Restart
  }

data Restart = Permanent | Transient | Temporary
data Strategy = OneForOne | OneForAll

data SupervisorSpec = SupervisorSpec
  { strategy  :: Strategy
  , intensity :: Int
  , period    :: Int
  , children  :: [ChildSpec]
  }

supervise :: SupervisorSpec -> ProcessM Pid
getChild  :: Pid -> String -> ProcessM (Maybe Pid)
\end{lstlisting}

The field \inlinehaskell{childStart} is an ordinary \inlinehaskell{ProcessM ()}.
It may be a generic server, another supervisor, or an application-specific process.
If \inlinehaskell{childNode} names a remote node, the child is spawned there.
If the child dies and the restart policy says it should be restarted, the supervisor simply spawns the same computation again.

For instance, a supervised remote counter server is just a child whose start action runs the generic server:

\begin{lstlisting}[style=mchhaskell]
  do let child = ChildSpec
           { childName    = "counter"
           , childStart   = runServer counterSpec
           , childNode    = Just n
           , childRestart = Permanent
           }

     sup <- supervise SupervisorSpec
       { strategy  = OneForOne
       , intensity = 3
       , period    = 5000
       , children  = [child]
       }

     Just pid <- getChild sup "counter"
     let counter = serverFromPid pid
     cast counter Inc
\end{lstlisting}

This is the role of the Erlang-style parts of \mch{} in this paper.
They are evidence that the direct \inlinehaskell{spawn} and
\inlinehaskell{send} interface is expressive enough to build structured
distributed programs, while keeping the central tradeoff visible:
process behavior is easy to move because it is represented as a runtime
graph, and the safety of that movement is correspondingly checked at
runtime.

\section{Implementation}
\label{sec:implementation}

\subsection{MicroHs \& Graph Reduction}
The Haskell implementation used in this paper is MicroHs~\cite{augustsson2024microhs}.
What matters for our purposes is not MicroHs itself, but the runtime support it provides: the serialization primitives that make the framework work.
The design of the framework is not tied to MicroHs, provided an alternative runtime exposes equivalent graph serialization primitives.

\paragraph{Serialization \& Deserialization}
MicroHs uses graph reduction with combinators~\cite{Turner1979}.  This means that a running program is
represented by a graph in memory that is continuously rewritten.  The leaf nodes in the graph are the combinators
and the interior nodes are application nodes.  There is sharing (which is necessary for lazy evaluation)
and cycles (because of recursion).
To serialize an arbitrary value, you simply have to traverse the sub-graph reachable from the starting node,
\textit{i.e.}, the value to serialize.

\paragraph{What can be serialized?}
In general, pure computations and values can be serialized without any problem.
If something cannot be serialized, the runtime will throw an exception.
Here is a more detailed breakdown of what works and what does not.

Can be serialized:
\begin{itemize}
\item application nodes
\item primitive operations, \textit{e.g.}, combinators, basic arithmetic, etc.
\item and by extension, any Haskell data type
\item basic values, \textit{e.g.}, fixed-size integers, floating point, etc.
\item byte strings
\item boxed arrays
\item \inlinehaskell{IORef}
\end{itemize}

Cannot be serialized:
\begin{itemize}
\item pointers to memory allocated with \inlinehaskell{malloc()}
\item foreign pointers
\item C function pointers
\item \inlinehaskell{MVar} values, which could be serialized, but are unlikely to work since threads are not serialized
\item thread IDs, which represent threads; we have chosen not to serialize these
\end{itemize}

\paragraph{Serialization API}
The underlying API exposed by the runtime system to serialize and deserialize values is simple:
\begin{lstlisting}[style=mchhaskell]
  serialize   :: a -> IO ByteString
  deserialize :: ByteString -> IO a
\end{lstlisting}

Taken in isolation, this API is unsafe.
\inlinehaskell{deserialize} makes no attempt to verify that the deserialized graph represents a value of the requested type.

\mch{} therefore treats these operations as low-level runtime primitives, not as the user-facing serialization API.
Their use is confined to protocol points where the runtime determines the expected shape of the graph.
For remote spawn, the payload is deserialized as the process function type expected by the node runtime.
For message delivery, a remote message is stored together with a \inlinehaskell{TypeRep}, and selective receive compares this representation before deserializing the message payload.

These checks do not make graph deserialization statically type safe.
They rely on compatible nodes following the \mch{} protocol and agreeing on the runtime graph format and type representations.
We return to richer boundary descriptors, such as a \inlinehaskell{TypeDescribable} constraint, in the Discussion.

\paragraph{Laziness}
Any graph can be serialized, regardless of how evaluated it is.  Thus unevaluated ``thunks'' can be serialized.
This is a feature, but it is also often not what you want.  It is easy to accidentally serialize a graph that was
actually intended to be evaluated first.
For data types, it is possible to control evaluation using \inlinehaskell{rnf} from the \inlinehaskell{NFData} class.
For functions, the situation is more delicate.
A function can only be forced to weak head normal form, which exposes a closure but does not force the values captured in its environment, nor any computation under the lambda.
Sending functions therefore preserves the direct style, but it does not remove the programmer's responsibility to understand what the closure reaches.
In practice, this means that values captured by mobile functions should be forced before the function is sent when their evaluation location matters.

\paragraph{Details of serialization}

The serialization is implemented as a two-pass algorithm.
The first pass identifies shared nodes via a depth-first traversal, whereas the second pass turns the graph into a byte string.
The information from the first pass is used to guide the second pass in where to reference previously serialized nodes rather than serializing them again.
Below follows an outline of the steps taken by the serializer algorithm:

\begin{itemize}
  \item First, two bit arrays are allocated, called \textit{visited} and \textit{shared}, with as many elements as there are nodes in the heap.
        These arrays are used as mark bits during the two passes.
        We use two arrays rather than reserving bits in the nodes themselves for two main reasons: fewer writes to memory, and the possibility of concurrent traversals.
  \item When the first pass traverses the graph and sees a node, a previously unvisited node will have its \textit{visited} bit set, whereas an already visited node will have its \textit{shared} bit set.
        Any outgoing edges from the node are then recursively visited.
  \item The second pass flattens the graph using a post-order traversal.
        If a node is both \textit{shared} and \textit{visited}, we unset its \textit{visited} bit, serialize the subparts, emit that node's textual representation, and finally emit a fresh label for that node.
        If a node is \textit{shared} and not \textit{visited}, it has already been serialized.
        In that case, we emit the label that was associated with the node when it was previously serialized.
        If a node is not \textit{shared}, it is serialized.
  \item Finally, the two temporary bit arrays are freed.
\end{itemize}
Labels can be represented in different ways.
The easiest representation is the address of the node, or perhaps the node's offset into the heap.
The representation does not matter as long as labels can be generated to be unique for each node.


\paragraph{Details of deserialization}
Deserialization is the inverse operation of serialization.
The textual format from the serializer is parsed, and an equivalent graph, preserving sharing and cycles, is reconstructed.
Below follows an outline of the steps taken by the deserializer algorithm:

\begin{itemize}
  \item First, we allocate an array that serves as a hash table, where labels are associated with node addresses.
  \item When a node is parsed, we check whether it is a label \textit{reference} or not.
        If it is, we look up the label in the hash table.
        The parser might see a label reference before the label \textit{definition}.
        In that case, we allocate a placeholder indirection node and later update it to point at the parsed definition, and return the address of the indirection node.
  \item After a node has been parsed, if we see a label \textit{definition}, we associate the label with the address of the just-parsed node in the hash table.
  \item Finally, we free the memory of the hash table and return the last-parsed node as the result of deserialization.
\end{itemize}

\paragraph{Format}
The serialization format is textual.
That makes it easier to read and debug, but it also avoids using fixed-width data for the basic types.

The format is portable across architectures in the sense that it does not expose machine byte order, pointer values, or binary code layout.
Portability depends on the sender and receiver using compatible runtime versions and on primitive values being representable on the receiving runtime.
In our current system, \inlinehaskell{Int} is an architecture-dependent primitive.
If a 64-bit runtime produces and serializes a value of type \inlinehaskell{Int}, the receiving runtime might not be able to represent the value if it is a 32-bit runtime.
Cross-architecture protocols that must rely on numeric ranges should use fixed-width types.
An optional compression/decompression pass can also be used to reduce the size of the serialized data.

\paragraph{Example}
Consider first the function \inlinehaskell{f} below:
\begin{lstlisting}[style=mchhaskell]
  f :: Int -> Int
  f x = x * x + 1
\end{lstlisting}
The serialized combinator representation of this function is:
\begin{lstlisting}[style=combformat]
v8.4
0
C' + @ S * @ I @@ #1 @
\end{lstlisting}
Here, \inlinecombformat{v8.4} is the format version, and \inlinecombformat{0} is the number of labels used.
This is followed by some combinators, with \inlinecombformat{@} being postfix application.
Another example illustrating a recursive function is shown below:

\noindent\begin{minipage}{\linewidth}
\begin{lstlisting}[style=mchhaskell,numbers=none]
fac :: Int -> Int
fac n = if n==0 then 1 else n*fac(n-1)
\end{lstlisting}
The serialized representation of this function is:
\begin{lstlisting}[style=combformat]
v8.4
1
C S C == @ #0 @@ S * @ B _49382130 @ C - @ #1 @@@@@#1 @ :49382130
\end{lstlisting}
Here, \inlinecombformat{_49382130} is a label reference and \inlinecombformat{:49382130} is a label definition (because the function is recursive).
\end{minipage}

\begin{figure}[t]
  \centering
  \begin{subfigure}[t]{0.38\linewidth}
    \centering
    \includegraphics[width=\linewidth]{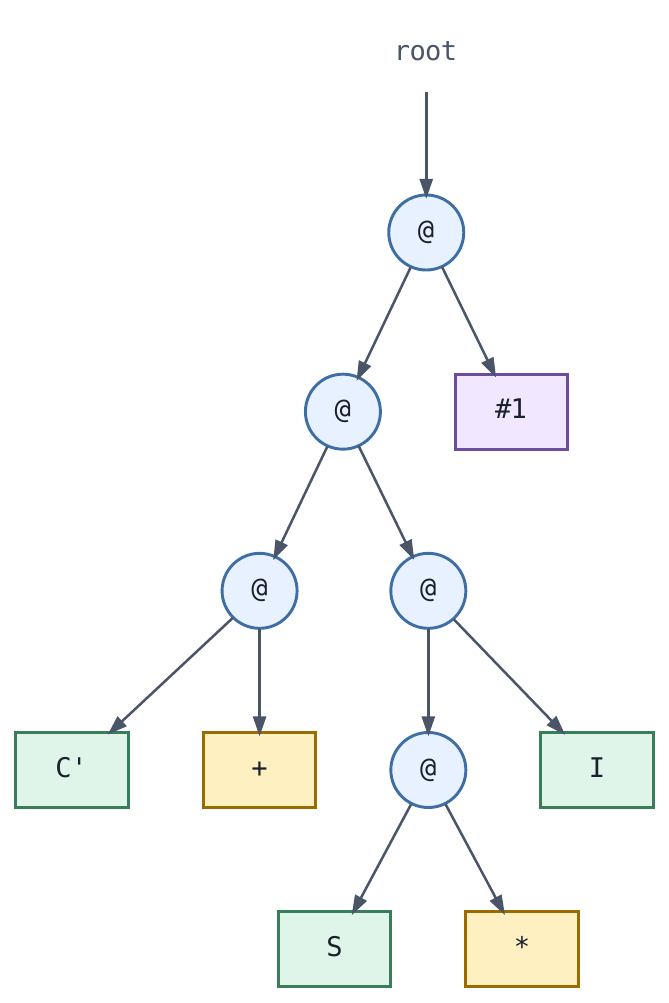}
    \caption{A non-recursive value.}
    \label{fig:combinator-graph-simple}
  \end{subfigure}
  \hfill
  \begin{subfigure}[t]{0.56\linewidth}
    \centering
    \includegraphics[width=\linewidth]{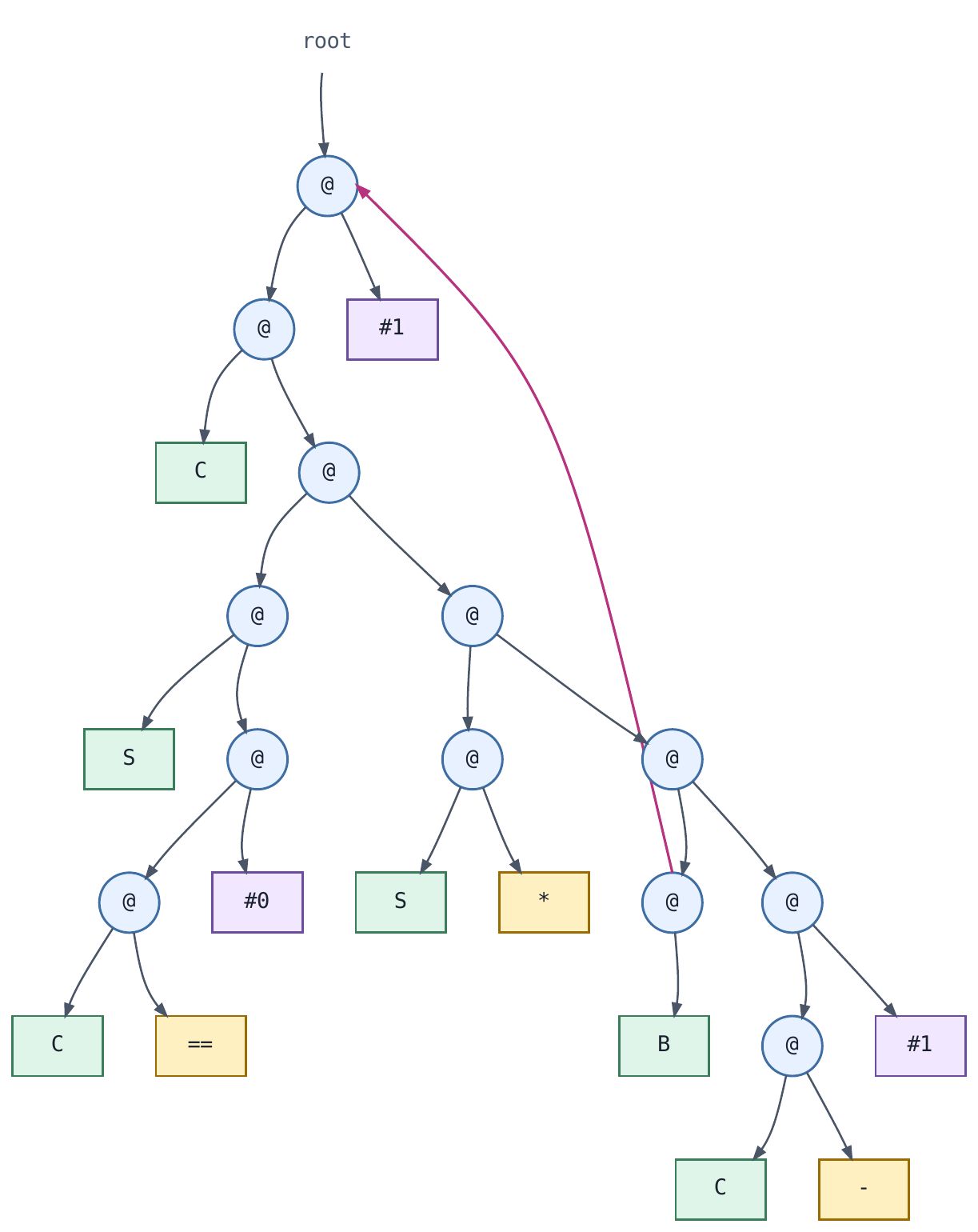}
    \caption{A recursive value.}
    \label{fig:combinator-graph-recursive}
  \end{subfigure}
  \caption{Combinator graphs corresponding to the two serialized examples.
  Application nodes are drawn as circles; combinators, primitive operations,
  and literals are drawn as leaves.  The recursive example requires a label
  in the textual serialization because the graph contains a cycle.
  The figure shows multiple copies of primitive nodes for clarity; in the runtime they are allocated once and shared.}
  \Description{Two combinator graph diagrams. The first is an acyclic graph
  for a simple arithmetic function. The second is a larger graph for factorial
  with a highlighted back edge showing recursion.}
  \label{fig:combinator-graphs}
\end{figure}

\Cref{fig:combinator-graphs} visualizes the two serialized examples above.
The purpose of the figure is simply to relate the flat textual format to the runtime object it describes: each \inlinecombformat{@} becomes an application node, while the other tokens become leaves in the graph.
In the factorial example, the label reference and label definition correspond to the recursive edge in the graph.
This graph, rather than a source-level closure name, is what \mch{} serializes when values or process bodies cross node boundaries.

Having described the graph serializer itself, we now describe where the distributed runtime invokes it: first in the node command loop, then in remote spawn and message delivery.

\subsection{Node Runtime Architecture}

A \mch{} program always runs a \textit{node}.
The node is invoked from \inlinehaskell{main} with a \inlinehaskell{NodeConfig}, a structure holding the node's identity (an IP address and port number).
The Haskell main thread creates an initial \textit{node state}, spawns a \textit{TCP listener} thread that awaits incoming connection requests, and then enters the main node loop.
After this initial setup, the node continuously awaits commands from either a remote peer over the network, or from a process running on the same machine.

This gives the runtime a simple shape, where processes and peer-reader threads typically do not manipulate the distributed runtime directly.
Instead, they communicate with the node loop by submitting commands.
The rest of this subsection explains that boundary: which requests can arrive at a node, where they come from, and how the node turns them into local runtime effects.

We first conclude that the node clearly needs to be able to process commands.
Examples of commands could be \textit{install this monitor}, \textit{send this message}, \textit{connect to this remote node}, to name just a few.
Some of the commands the node needs to process come from processes that are running on the node in question, whereas some commands will be sent from remote nodes over the network.
We model the commands without much flair:

\begin{lstlisting}[style=mchhaskell]
  -- | A command for the node to handle
  data NodeCommand
    = Local Command          -- ^ A command from a process on this node
    | NonLocal PeerCommand -- ^ A command sent from another node

  -- | Commands that a local process sends to the node
  data Command where
    Connect :: SockAddr -> Maybe (MVar ()) -> Command
    Deliver :: Pid -> Mail -> Command
    -- more

  -- | A command that a remote node has asked the node to execute
  data PeerCommand
    = PeerConnect SockAddr Socket [SockAddr]
    | PeerDeliver Pid Mail
    -- more

  \end{lstlisting}

The node processes commands as they arrive in a \inlinehaskell{cmdQueue :: CommandQueue} maintained in its node state.
The \inlinehaskell{CommandQueue} is, internally, a double-ended queue coupled with an \inlinehaskell{MVar} used to signal when an empty queue has been written to.
\begin{lstlisting}[style=mchhaskell]
  data Queue a = Queue [a] [a]
  data CommandQueue a = CommandQueue (MVar (Queue a)) (MVar ())

  newCommandQueue :: IO (CommandQueue a)
  enqueueCommand :: CommandQueue a -> a -> IO ()
  takeCommand :: CommandQueue a -> IO a
\end{lstlisting}
A more robust implementation might assign different priorities to commands, but our implementation treats them all as equally important.

A local process will construct a \inlinehaskell{Local Command} and write it to the command queue, whereas a remote command will be received by a green thread monitoring a socket, as a \inlinehaskell{NonLocal PeerCommand}.
While a command is traveling over the network, it is represented as yet another type of command, \inlinehaskell{WireCommand}.

\begin{lstlisting}[style=mchhaskell]
  data WireCommand
    = WireConnect SockAddr [SockAddr]
    | WireDeliver Pid Int Int
    -- more
\end{lstlisting}

During transit, a command will have been turned primarily into a buffer of bytes.
In the example constructors above, the \inlinehaskell{WireDeliver} variant carries a process identifier and two sizes, denoting the length of the two following data packets (a serialized \inlinehaskell{TypeRep}, and the \inlinehaskell{Mail} itself).
The TCP listener will convert a \inlinehaskell{WireCommand} to a \inlinehaskell{PeerCommand} by, \textit{e.g.}, reading potential buffers and reconstructing values that have been serialized.

\begin{figure}
  \centering
  \includegraphics[scale=0.5]{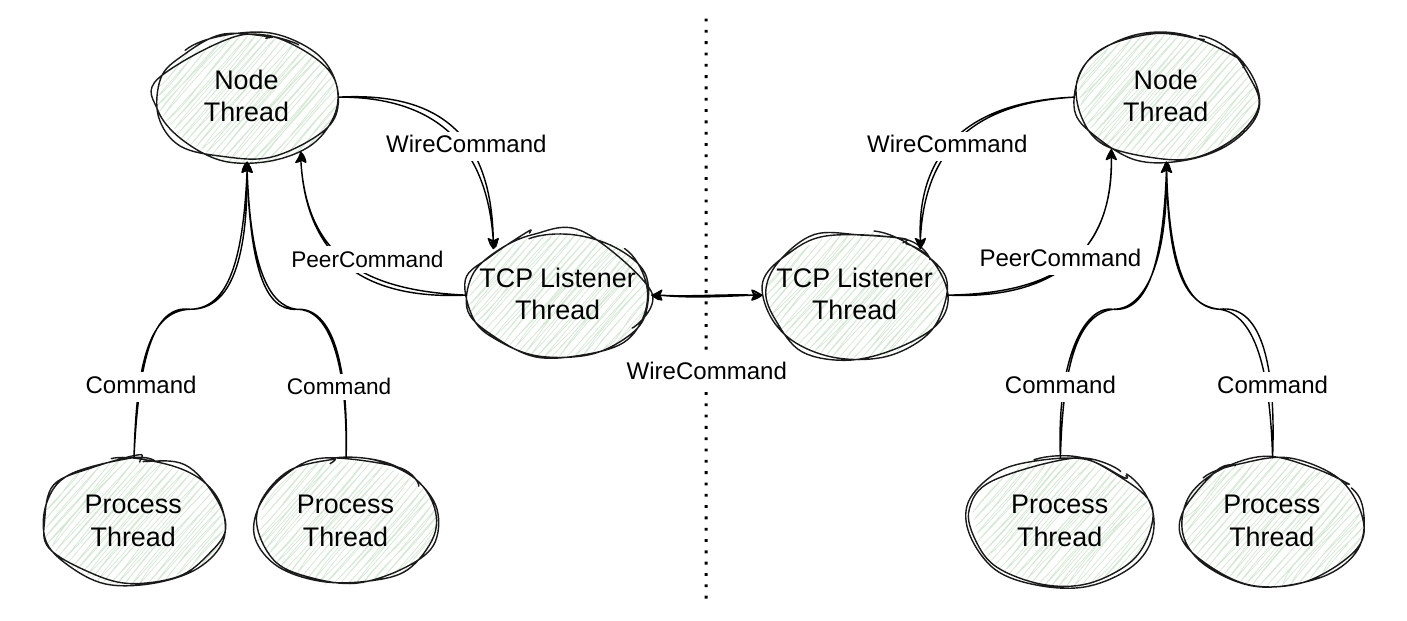}
  \caption{The figure illustrates how the different types of commands flow through the node application. Every node in this diagram is a Haskell green thread, and the dotted line represents the network boundary. Process threads will send the node ordinary \inlinehaskell{Command} values, whereas the node thread needs to convert such a command to an appropriate \inlinehaskell{WireCommand} before the TCP listener thread can send it to a remote node. A TCP listener converts a received \inlinehaskell{WireCommand} to a \inlinehaskell{PeerCommand} before sending it to the node thread.}
  \label{fig:commands}
\end{figure}

The node processes commands in the order in which they arrive.
A figure that shows how the different types of commands flow through the node application is shown in \cref{fig:commands}.
The command interface is what lets the node remain the owner of the runtime state.
Mailboxes, monitors, connections, pending receives, and the name registry are not exposed to processes directly, but are rather updated only by the node loop while handling commands.
Instead of showing the entire node state definition at once, we will introduce the relevant pieces as they become necessary.

\subsection{The ProcessM Monad}

The \inlinehaskell{ProcessM} monad is a central part of \mch{}, as it models a \textit{process}.
Conceptually, a process is an independent entity that performs computations regardless of what other processes do, and communicates with other processes only via message passing.
It is modelled as a function that takes the node state and its own pid, and then produces an \inlinehaskell{IO} action.

\begin{lstlisting}[style=mchhaskell]
data ProcessM a = ProcessM { unProcessM :: NodeState -> Pid -> IO a }
\end{lstlisting}

The process needs to know about the node state as the node state supplies---most notably---the \inlinehaskell{MVar} through which commands are sent.
A common pattern in \inlinehaskell{ProcessM} primitives that must wait for a result from the node is that they allocate a fresh \inlinehaskell{MVar}, send its command to the node (which includes the \inlinehaskell{MVar}), and then block until the result is there.
An example is shown below:

\begin{lstlisting}[style=mchhaskell]
  -- | Register a Pid under a name
  register :: Pid -> String -> ProcessM ()
  register pid name = ProcessM $ \ns _ -> do
    -- create a fresh MVar to block until the node has registered the name
    mv <- newEmptyMVar

    -- send the command to the node
    enqueueCommand (cmdQueue ns) (Local (Register pid name mv))
    
    -- wait for the node to write to the MVar, indicating that it is done
    takeMVar mv
\end{lstlisting}

Additionally, functions that return a result fetched from the node state must force evaluation of it.
Consider the function \inlinehaskell{node} that returns the node on which the process is currently executing.

\begin{lstlisting}[style=mchhaskell]
  node :: ProcessM NodeId
  node = ProcessM $ \ns _ ->
    let nd = NodeId $ nodeAddr ns
    in rnf nd `seq` return nd
\end{lstlisting}

If we did not force the evaluation of the \inlinehaskell{NodeId}, there is a high risk that the result from \inlinehaskell{node} would be an unevaluated thunk.
We illustrate this problem with the code snippet below, which shows a process that spawns another process that refers to the fetched \inlinehaskell{NodeId}.

\begin{lstlisting}[style=mchhaskell]
do n <- node
   ns <- filter ((/=) n) <$> nodes
   mapM_ (\remoteNode -> spawn remoteNode (spawn n (liftIO $ putStrLn "hi"))) ns
\end{lstlisting}

The above program fetches the current process's node, and then spawns a process on every other remote node.
That process in turn spawns a process on the original node again, by referring to the free variable \inlinehaskell{n}.
If \inlinehaskell{n} was an unevaluated thunk that refers to the node state, the outer call to \inlinehaskell{spawn} would not succeed because the serializer would follow the graph into the node state and try to serialize its inner \inlinehaskell{MVar}s.
By ensuring the result is forced, we know that the node state is no longer referenced by \inlinehaskell{n} and serializing proceeds as expected.

This is also where the conceptual process and the runtime bookkeeping part ways.
A \inlinehaskell{ProcessM} computation is given a \inlinehaskell{Pid}, but the resources associated with that pid are owned by the node.
The process has a mailbox, may have pending receives, may be registered under a name, and may be monitored by other processes.
However, none of these are fields of the process itself.
They are entries in the node state, and the process can affect them only by sending commands to the node.

This distinction is important for distribution.
When a process is spawned remotely, we do not serialize its mailbox, monitors, or other local runtime resources.
We serialize only the computation to run.
The receiving node allocates a fresh pid, creates the corresponding local runtime entries, and then runs the computation with its own node state.

\subsection{Remote Spawn}

The central contribution of \mch{} is that \inlinehaskell{spawn} can take an ordinary \inlinehaskell{ProcessM} computation and run it on another node.
We now describe how a call to \inlinehaskell{spawn} moves through the node runtime, crosses the network, and results in a fresh green thread running the supplied computation on a remote machine.
The example program we begin with is shown below:

\begin{lstlisting}[style=mchhaskell]
  do n <- node
     anotherNode <- (head . filter ((/=) n)) <$> nodes

     let x = 7

     _ <- spawn anotherNode $ do
            liftIO $ putStrLn $ show x
            let x = 10
            liftIO $ putStrLn $ show x
  
     return ()
  \end{lstlisting}

We note that the spawned process refers to the outer \inlinehaskell{x}, a free variable.

The implementation of \inlinehaskell{spawn} has a ``fast path'' for node-local spawns, to achieve better performance.
The rest of this section, however, describes how a remote spawn is implemented (the ``slow path'').
The fast path differs from the slow path in that the running process spawns the new process itself, bypassing the node loop/command queue.
Additionally, the process body of a locally spawned process is not serialized, as the process body does not need to traverse the network.
The slow path presented below describes a superset of the actions that the fast path takes.

As \inlinehaskell{spawn} returns the \inlinehaskell{Pid} of the spawned process, it blocks until the remote node has created the process and acknowledged its \inlinehaskell{Pid}.
A fresh \inlinehaskell{MVar} is allocated where the node will write the \inlinehaskell{Pid} once it receives it, along with a spawn confirmation from the remote node.
Additionally, a unique token is generated and paired with the spawn request, so that an acknowledgment of another spawn request is not confused with the current request.
The result \inlinehaskell{MVar} is associated with the unique token in a mutable map reachable from the node state.

\begin{lstlisting}[style=mchhaskell]
spawn :: NodeId -> ProcessM () -> ProcessM Pid
spawn (NodeId targetAddr) action = ProcessM $ \ns _myPid -> do
  -- generate token and create MVar, to record in the node state
  reqId    <- newUniqueToken
  replyVar <- newEmptyMVar
  modifyIORef' (spawnRequests ns) (Map.insert reqId replyVar)

  -- send the spawn command to the node loop
  enqueueCommand (cmdQueue ns) (SpawnOn targetAddr reqId (unProcessM action))

  -- block here until the node loop publishes the process identifier of the
  -- spawned process
  pid <- takeMVar replyVar
  modifyIORef' (spawnRequests ns) (Map.delete reqId)

  return pid
\end{lstlisting}

When the node processes the \inlinehaskell{SpawnOn} command, it fetches the socket associated with the remote node and initiates the network communication.

\begin{lstlisting}[style=mchhaskell]
  -- this function pattern matches on many commands, but we just show SpawnOn here
  handleLocalCommand :: NodeState -> Command -> IO ()
  handleLocalCommand ns (SpawnOn targetAddr reqId f) = do
    c <- getConn ns targetAddr
    sendSpawn c (nodeAddr ns) reqId f

  -- remember that the ProcessM monad wrapped a function of type
  -- NodeState -> Pid -> IO a
  sendSpawn :: Socket -> SockAddr -> Int -> (NodeState -> Pid -> IO ()) -> IO ()
  sendSpawn conn spawnerAddr reqId f = do
    (buf, len) <- enc f -- serialize the ProcessM function

    -- send first the serialized WireSpawn command, and then the serialized graph
    sendLine conn (show (WireSpawn spawnerAddr reqId len))
    sendAll conn buf len

    free buf
\end{lstlisting}

Here is where we invoke our magic serialization function.
The serialized graph will reference everything that was reachable from the expression, including the outer free variable.
This is now a serialized function that contains both its code and environment.

After serialization, we get a pointer to a memory buffer holding our ready-to-send bytes.
The \inlinehaskell{WireCommand} that we send to the remote node includes the unique token that we have associated with the spawn request, as well as the length of the following data buffer.
The remote node's TCP listener will first receive the \inlinehaskell{WireCommand}, look at it to figure out how many bytes to read for the subsequent graph, and then decode it when it turns it into a \inlinehaskell{PeerCommand}.
We omit most of the listener code, but show how the received \inlinehaskell{WireSpawn} message is treated:

\begin{lstlisting}[style=mchhaskell]
  handleWireCommand (WireSpawn spawnerAddr reqId payloadLen) = do
    buf <- mallocBytes payloadLen
    ok  <- recvExact conn buf payloadLen

    if not ok
      then do
        -- error handling code, omitted for brevity
      else do
        f <- dec buf payloadLen :: IO (NodeState -> Pid -> IO ())
        free buf
        enqueueCommand (cmdQueue ns) (NonLocal (PeerSpawn spawnerAddr reqId f))
\end{lstlisting}

After the \inlinehaskell{WireSpawn} command has already been parsed, we know that we should read a subsequent buffer of data, the graph.
We allocate a buffer of the appropriate length and read exactly that many bytes into it.
We then invoke our magic deserializer, which reconstructs the graph on the heap of the remote \mch{} node.
Here we annotate the deserializer with the type of the function that \inlinehaskell{ProcessM} wraps.
There is no verification that the parsed graph represents a value of that specific type, so we must be very careful to properly line up the types of the values we serialize, and the values we deserialize.
We then enqueue the \inlinehaskell{PeerSpawn} command with the node loop.

The node loop on the remote node must invoke its \inlinehaskell{PeerCommand} handler to deal with the command.
It starts the local process, and sends the \inlinehaskell{Pid} along with the unique token back to the original node.
We omit the code for \inlinehaskell{startLocalProcess}, but mention that it uses \inlinehaskell{forkIO} to spawn a green thread that runs the \inlinehaskell{ProcessM} function we just deserialized.
The process's mailbox and associated metadata are then added to the node state.

\begin{lstlisting}[style=mchhaskell]
  handleNonLocalCommand ns (PeerSpawn spawnerAddr reqId f) = do
    pid <- startLocalProcess ns f
    c   <- getConn ns spawnerAddr
    sendLine c (show (WireSpawnAck reqId pid))
\end{lstlisting}

At the wire level, the original node receives the \inlinehaskell{WireSpawnAck}, which the listener promptly converts into a \inlinehaskell{PeerSpawnAck}.
The spawn is then completed by invoking the
\mbox{\inlinehaskell{completeSpawnRequest}}
function that was briefly shown in the code that described how the node handled the initial spawn request.
The function \inlinehaskell{completeSpawnRequest} publishes the \inlinehaskell{Pid} into the \inlinehaskell{MVar} that is associated with the unique token.
This unblocks the process that initiated the spawn request, which now has access to the process identifier of the new process.
This concludes the handling of \inlinehaskell{spawn}.

\subsection{Message Passing}

The underlying mechanism for transmitting messages between communicating processes is the same.
There is a fast and a slow path, differing in exactly the same way that the fast and slow paths did for \inlinehaskell{spawn}.
Because values, like code, are just subgraphs in the combinator graph, the serializer and deserializer work here as well.
If a message has to traverse the network it will be encoded together with a runtime representation of its type.
If a message does not have to traverse the network, and is being sent to a process running on the same node, the message is passed by its pointer into the graph.
We capture these two kinds of messages in the following data type:

\begin{lstlisting}[style=mchhaskell]
  data Mail where
    HeapMail    :: a -> Mail
    EncodedMail :: TypeRep -> Ptr Word8 -> Int -> Mail
\end{lstlisting}

We use \inlinehaskell{TypeRep} from the \inlinehaskell{Typeable} type class to observe the types of values at runtime.
Both GHC and MicroHs derive \inlinehaskell{Typeable} instances for every type, so the programmer does not need to implement them.

Every process has a mailbox where its messages go, but the mailbox is owned by the node loop.
In the fast path, the message will be delivered directly to the receiving process, whereas in the slow path, the action of sending the message will be delegated to the node.
We omit the code for \inlinehaskell{deliverMailLocal}, which is responsible for doing the actions just described.
The code for \inlinehaskell{send} itself is fairly brief:

\begin{lstlisting}[style=mchhaskell]
  send :: Identifiable process => process -> a -> ProcessM ()
  send targetPid x = ProcessM $ \ns _ -> do
    targetPid' <- toPid targetPid ns
    let Pid (NodeId targetAddr) _ = targetPid'
    if targetAddr == nodeAddr ns
      then deliverMailLocal Nothing ns targetPid' (HeapMail x)
      else
        enqueueCommand (cmdQueue ns) (Local (Deliver targetPid' (HeapMail x)))
\end{lstlisting}

There are two types that are \inlinehaskell{Identifiable}: \inlinehaskell{Pid} and \inlinehaskell{String}. 
The function \inlinehaskell{toPid} does a lookup in an \inlinehaskell{IORef} fetched from the node state if the instantiated type is \inlinehaskell{String}.

When the node handles the remote send, it will fetch the associated socket, encode both the message and the \inlinehaskell{TypeRep}, and then send everything through the socket.

\begin{lstlisting}[style=mchhaskell]
  handleLocalCommand ns (Deliver targetPid mail) = do
    let Pid (NodeId targetAddr) _ = targetPid
    deliverMailRemote ns targetPid mail

  deliverMailRemote :: NodeState -> Pid -> Mail -> IO ()
  deliverMailRemote ns targetPid (HeapMail x) = do
    let Pid (NodeId targetAddr) _ = targetPid
    c <- getConn ns targetAddr
    
    (valueBuf, valueLen) <- enc x
    (trBuf, trLen)       <- encodeTypeRep (typeOf x)
    
    sendDeliver c targetPid trLen trBuf valueLen valueBuf
    
    free valueBuf
    free trBuf
  deliverMailRemote _ _ (EncodedMail _ _ _) = -- should not happen
\end{lstlisting}

When a message arrives for a process, the thread handling its delivery checks whether the process is blocked waiting to receive a message.
If so, the delivering thread tests the message against the pending receive.
A matching message is passed directly to the blocked receive, bypassing the mailbox; a non-matching message is placed in the mailbox without waking the process.
This avoids waking the process only to have it reject the message and block again.

\paragraph{Forcing values before sending}

Something to keep in mind when sending messages between processes is that messages will be sent as they are, then and there.
The serializer walks the combinator graph, and if a value is still an unevaluated thunk, it will be serialized and transmitted.
This might be the intended behavior, but it will most likely be the intent of the programmer that an evaluated value should be sent.
The problems with sending unintended unevaluated thunks are twofold: firstly, a seemingly small expression might make a large part of the sender's heap be serialized and sent to the remote node;
secondly, if we are using processes and message passing to achieve speedups by distributing our computation over a network, work might not be performed on those nodes where the programmer expects it to be done.
We note that in \ch{}, this concern does not arise.
The \inlinehaskell{Binary} instance required by \ch{}'s \inlinehaskell{send} requires a value to be evaluated as it is serialized.

To give an example of misplacement of work, consider \inlinehaskell{countChunk} below.
It takes a list of numbers as input and sends a message to the process identifier indicating how many of those numbers were prime.
Since \inlinehaskell{n} is not forced, its thunk will point to \inlinehaskell{xs}, \inlinehaskell{filter}, \inlinehaskell{length}, and \inlinehaskell{isPrime}.
Serializing \inlinehaskell{n} will serialize those functions as well, and when \inlinehaskell{n} is required at the parent, the work will be done there.

\begin{lstlisting}[style=mchhaskell]
  countChunk :: Pid -> [Int] -> ProcessM ()
  countChunk parent xs = do
    let n = length (filter isPrime xs)
    send parent n
\end{lstlisting}

The fix is to ensure that \inlinehaskell{n} is forced before it is given to \inlinehaskell{send}.

\begin{lstlisting}[style=mchhaskell]
  countChunkStrict :: Pid -> [Int] -> ProcessM ()
  countChunkStrict parent xs = do
    let n = length (filter isPrime xs)
    n `seq` send parent n
\end{lstlisting}

In this case, we can use \inlinehaskell{seq} to force evaluation, since \inlinehaskell{n} is an \inlinehaskell{Int}, but more structured data would require \inlinehaskell{rnf} or \inlinehaskell{deepseq}.

This is the same discipline familiar to programmers of parallel Haskell.
A spark created by \inlinehaskell{par} only buys parallelism if the sparked expression is itself evaluated by another core.
In \mch{}, the analogous failure mode is not lost parallelism, but misplaced computation, and possibly a much larger message than the programmer intended.
The remedy is the same: ensure the message is in normal form before handing it to \inlinehaskell{send}, using \inlinehaskell{seq}, bang patterns, or a deep-force such as \inlinehaskell{deepseq}.

\subsection{Failure Detection and Exit Propagation}

The fault-tolerance primitives in \mch{} are built around process-death detection.
The runtime does not try to recover a failed process by itself, but instead makes failure observable to other processes.
Application-level code, such as supervisors, can be programmed to decide whether a process should restart or not.

As described previously, one process can observe another's death by installing a monitor.
Because the node owns the entire state, it maintains a table which maps each watched process to the processes that in turn watch it, and the monitor action each watcher has requested.

\begin{lstlisting}[style=mchhaskell]
  monitors :: IORef (Map.Map Pid [(Pid, MonitorAction)])
\end{lstlisting}

Unsurprisingly, a monitor is registered by sending a command to the node loop.
If the process that should be watched is local, the node records the monitor in its \inlinehaskell{monitors} table.
Instead, if the process lives on another node, the request is sent to the remote node as a \inlinehaskell{WireMonitor} command.
The remote node will record the monitor in its own \inlinehaskell{monitors} table.
The short summary is that a node is responsible for detecting process death and recording which watchers must be notified.

Every \mch{} process is implemented by a Haskell green thread.
When a process is started, the runtime forks the thread with a finalizer.
This finalizer runs regardless of whether the process returns normally, calls \inlinehaskell{terminate}, dies because of an exception, or is killed by another process via \inlinehaskell{exit}.

\begin{lstlisting}[style=mchhaskell]
  startLocalProcess :: NodeState -> String -> (NodeState -> Pid -> IO ()) -> IO Pid
  startLocalProcess ns label f = do
    pid <- allocatePid ns
    tid <- forkFinally (f ns pid) $ \res -> do
      let reason = reasonFromResult res
      cleanupProcess ns pid
      notifyMonitors ns pid reason
    modifyIORef' (threadIds ns) (Map.insert pid tid)
    return pid
\end{lstlisting}

The finalizer turns the result of the thread into an \inlinehaskell{ExitReason}.
Normal return and \inlinehaskell{terminate} become \inlinehaskell{ExitNormal}, whereas termination by \inlinehaskell{exit} preserves the \inlinehaskell{ExitReason} supplied by the caller.
An uncaught exception becomes \inlinehaskell{ExitOther}.
The runtime will then remove the process-local state such as its mailbox, pending receives, registered name, and thread identifier, before finally notifying the monitors.

Notification is also routed through the node loop.
For each registered watcher, the node loop checks whether it is a local process or not.
A local watcher is notified immediately, whereas a remote watcher is notified by sending a \inlinehaskell{WireProcessDied} command to the watcher's node.

\begin{lstlisting}[style=mchhaskell]
  applyMonitorAction ns watcherPid watcheePid TrapExit reason =
    deliverMailLocal ns watcherPid (HeapMail (ProcessDied watcheePid reason))

  applyMonitorAction ns watcherPid watcheePid Succumb _ = do
    tids <- readIORef (threadIds ns)
    case Map.lookup watcherPid tids of
      Just tid -> throwTo tid (MonitoredProcessDied watcheePid)
      Nothing  -> return ()
\end{lstlisting}

A monitor registered with \inlinehaskell{TrapExit} will receive the notification via a \inlinehaskell{ProcessDied} message in its mailbox.
The monitor process can then observe the termination and decide what to do.
Supervisor processes use this behavior to figure out whether a process should be restarted or not.
The \inlinehaskell{Succumb} monitor action will make the monitor receive an asynchronous exception, causing it to die together with the watched process.
This is useful to tie together the lifetimes of processes that are meant to exist at the same time.

Node disconnection is treated as process failure.
If an open socket is closed, the surviving nodes will assume that any monitored process on that node has terminated.
The surviving nodes will consider any monitored process as terminated with an \inlinehaskell{ExitOther} reason.
This makes the running processes able to detect network failure as if it were ordinary process failure.

Explicit \inlinehaskell{exit} signals use the same path.
If the target process is running on the current node, the node raises a \inlinehaskell{ProcessExit} exception in its thread, whereas if the process is remote, the node instead sends a \inlinehaskell{WireExit} command to the node that owns the process.
In either case, the target process eventually runs the same finalizer.

\section{Evaluation}

We evaluate \mch{} as a prototype implementation of a design point, rather than as a mature high-performance actor runtime.
The evaluation therefore asks four questions.
First, what are the basic costs of process creation, message passing, and runtime graph serialization?
Second, can a distributed \mch{} program obtain a speedup on a parallel workload despite those costs?
Third, is the programming model expressive enough to structure a nontrivial distributed application?
Finally, does the small MicroHs runtime make the same model usable on platforms outside the reach of a conventional GHC-based system?

The ring benchmark addresses the first question, while the distributed work-pool benchmark addresses the second.
The \ms{} case study addresses expressiveness by exercising named processes, dynamic process creation, generic servers, supervisors, monitors, and cross-node messaging in one application.
The heterogeneous-node case study addresses portability by running \mch{} nodes across microcontrollers and an ordinary laptop.

\subsection{Benchmarks}

\paragraph{Ring Benchmark}
\label{subsec:paragraph:ring-benchmark}

We take inspiration from Erlang with our first evaluation.
A classic Erlang example is to spawn a \textit{ring} of processes, where a message is sent from one process to another, through the entire ring, until the message arrives again at the original process.
We use the ring of processes to measure how long it takes, on average, to spawn a single worker, and how long it takes to send a message from one process to another.
We build a ring of 3000 processes.

All processes in the ring are spawned on the same node.
Node-local \inlinehaskell{spawn} and \inlinehaskell{send} have fast-paths whereby the actions are performed directly by the process rather than the node, avoiding the need to hand over to the node thread.
We can, however, instrument \mch{} to take the long path through the node loop, simulating what would happen if the \inlinehaskell{spawn} or \inlinehaskell{send} targeted another node.
Additionally, even for node-local \inlinehaskell{spawn} or \inlinehaskell{send}, we can force serialization and deserialization, to measure how long those operations take.
We measure the performance both with and without serialization enabled, to gain some insight into how much time encoding and decoding the graph takes.
The serialized process is roughly 32 kilobytes of data, whereas the serialized message sent through the ring is roughly 300 bytes.

We run this experiment on a laptop with an Intel i5-8365U CPU, and 16 GB of RAM.

\begin{figure}
  \centering
  \includegraphics[scale=0.15]{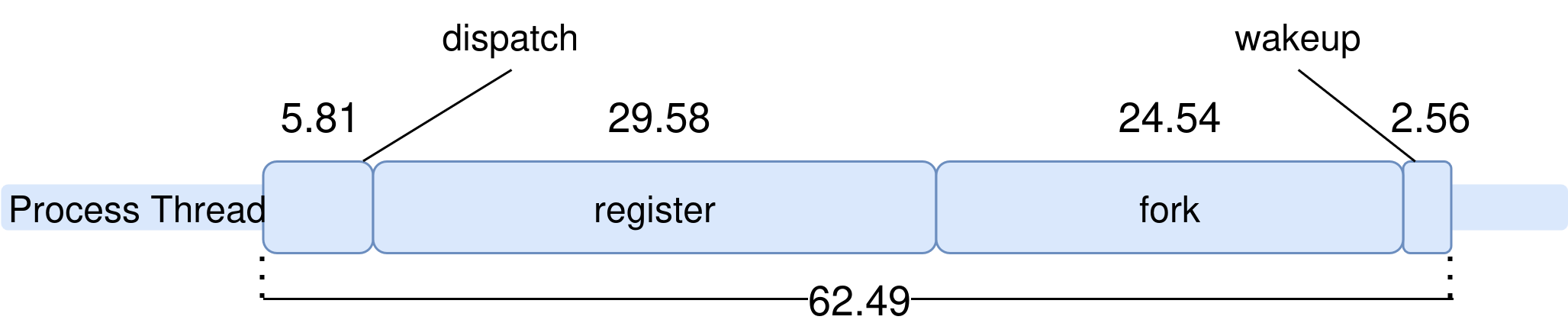}
  \caption{Breakdown of node-local process spawn time on the fast path, where the spawning process creates the child directly without routing through the node loop.}
  \label{fig:nodelocalspawn}

  \vspace{1em}

  \centering
  \includegraphics[scale=0.15]{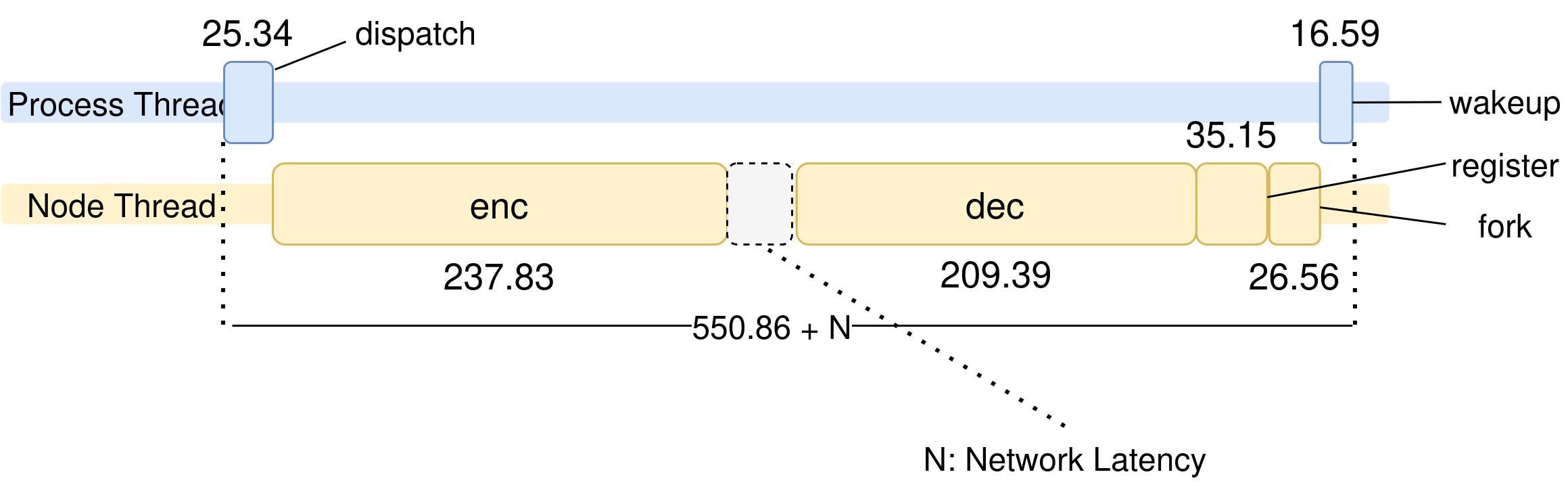}
  \caption{Breakdown of process spawn time on the slow path, where the process body is serialized, routed through the node loop, deserialized, and then started locally. A remote spawn on comparable hardware would add the network round-trip and transmission time.}
  \label{fig:nodelocalspawnslowpath}
\end{figure}

\Cref{fig:nodelocalspawn} and \cref{fig:nodelocalspawnslowpath} present the elapsed time when spawning a process on the same node, using either the fast path or the slow path.
The fast path takes roughly 60 \si{\micro\second}, whereas the slow path takes, for this example, 550 \si{\micro\second}.
The elapsed time is reported as segments, to try to measure what it is precisely that is taking time.
The reported segments are:

\begin{itemize}
  \item \textbf{dispatch} -- The running process handing off to the thread responsible for doing the \inlinehaskell{spawn}. In the fast path, this is roughly figuring out that the spawning process itself should do it, whereas in the slow path, it is a context switch to the node thread.
  \item \textbf{register} -- The time it takes to create and initialize process resources, such as mailboxes and the process identifier, and to register them in the node state.
  \item \textbf{fork} -- The time it takes to actually spawn the green thread that will execute the new process. Additionally, the node state is updated to include the fresh thread ID of the spawned thread.
  \item \textbf{wakeup} -- The time it takes to hand control back to the initial process, which is done by writing the fresh process identifier to an \inlinehaskell{MVar} that the initial process reads from.
  \item \textbf{enc} -- The time the runtime takes to serialize the reachable subgraph representing the process body to be spawned.
  \item \textbf{dec} -- The time it takes for the runtime to deserialize and reconstruct the graph from the serialized representation.
\end{itemize}

It takes \mch{} 371,347 \si{\micro\second} to send a message through the entire ring.

For comparison, a state-of-the-art actor model concurrency system, Erlang, takes on average 1.27 \si{\micro\second} to spawn a single process.
Sending a message through a ring of 3000 processes takes on average 1456 \si{\micro\second}.

\paragraph{Distributed Work Pool Benchmark}

This benchmark measures whether adding \mch{} nodes yields proportional speedup on an embarrassingly parallel workload.
We count the primes in a fixed integer interval, dividing it into equally sized segments that a work pool distributes as units of work across remote workers.
The work pool is implemented as a \mch{} application, where a master process takes N units of work, spawns M worker processes, and distributes the units of work to the workers one at a time.
A unit of work is modelled as a \inlinehaskell{ProcessM T} computation.

\begin{figure}
  \centering
  \includegraphics[scale=0.5]{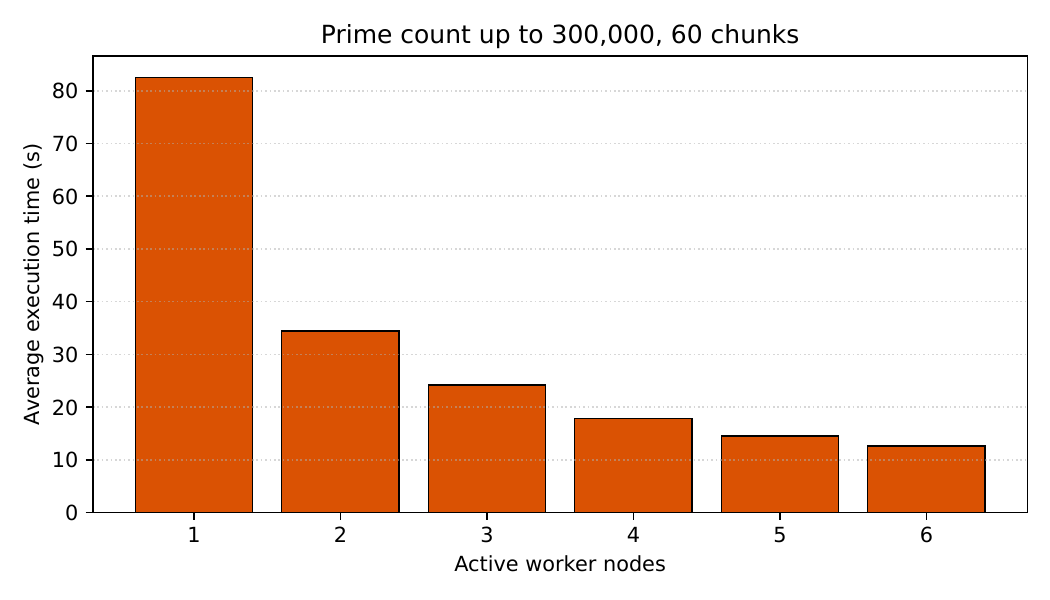}
  \caption{Distributed work-pool benchmark for counting primes up to 300{,}000, split into 60 work units. Bars show mean execution time over 10 runs on one to six Google Cloud E2-standard-2 worker nodes; runtime drops from 82.5\,s to 12.6\,s, a 6.5$\times$ speedup over the one-node baseline.}
  \label{fig:workpool-results}
\end{figure}

We distribute the workload over 6 remote Google Cloud E2-standard-2 machines, each equipped with 8 GB of RAM and 2 vCPUs.
Each instance runs a \mch{} node; these nodes are connected together.
We run the workload using between one and six nodes, averaging over 10 runs.
The results can be seen in \cref{fig:workpool-results}.

\subsection{Case Study 1: MicroSync}

To illustrate the programming model beyond small code snippets and toy examples, we implement a file syncing application, \ms{}.
The case study is modest in scope, but deliberately nontrivial.
Several machines work together to keep the contents of a directory in sync, discover each other, broadcast local edits, exchange metadata, exchange partial chunks of files, and recover from failing helper processes.
The case study shows that this application is expressed fairly naturally as a collection of communicating processes.

\begin{figure}
  \centering
  \includegraphics[scale=0.5]{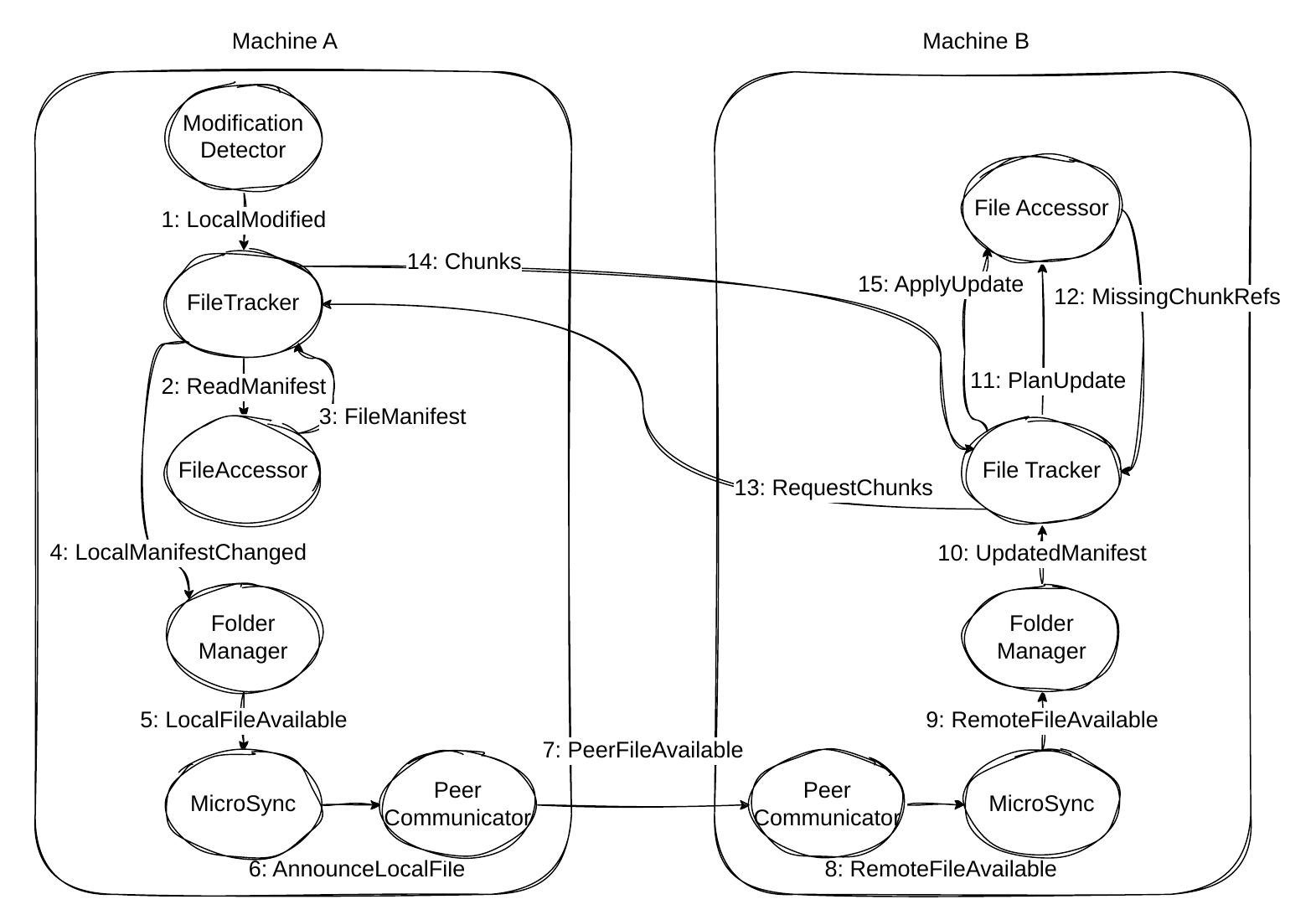}
  \caption{Message flow for synchronizing a local edit from machine A to machine B. The numbered arrows show the messages described in the walkthrough: the update is first advertised through the peer processes, after which the receiving file tracker requests missing chunks directly from the file tracker that advertised the version.}
  \label{fig:sync-flow}
\end{figure}

We dive straight in, and illustrate how an edit of a local file on machine A is advertised and applied to a remote copy of the same file on machine B; \cref{fig:sync-flow} shows the corresponding message flow.

On machine A, a user modifies a local file.
The \textit{modify detector} process assigned to monitor that particular file detects this edit by observing the last modified timestamp, and reports this fact to its parent \textit{file tracker} process.
The file tracker process is \ms{}'s designated owner of one tracked file.
It receives the message from the modification detector that the file has likely been modified, and asks the \textit{file accessor} for the file manifest.
A file accessor process is the only process that is allowed to modify a file on the disk, and is implemented as a server.

A file manifest is a compact description of a file version. It records, among other things, the path to the file, the file size, and most importantly, a list of chunk references.
Each chunk reference contains an offset into the file, the size of the chunk, and a hash of the chunk contents.
\ms{} uses \textit{Content-Defined Chunking} to divide a file into chunks.
The chunks are chosen based on the contents of the file, rather than using fixed byte offsets.
The effect of this is that a small edit tends to change only chunks close to where the edit happened, allowing a remote \ms{} instance to request only the chunks that it does not already have.

Once the file accessor has supplied the file manifest, the file tracker sends it to the \textit{folder manager} process.
The folder manager process is the designated owner of the entire folder being synchronized.
For this local-modification case, its main role is to turn the tracker's manifest into a file-version announcement that is sent upward to the top-level \textit{\ms{}} process.
The top-level \ms{} process routes the announcement to its \textit{peer communicator} process.
The peer communicator process informs the remote peer communicator on machine B what machine A's version of the file looks like.
On machine B, the information flows in a similar way from the peer to the \ms{} top-level process, from there to the folder manager, which will in turn inform the responsible file tracker process that there exists a potentially new version of the file.

The file tracker on machine B will ask its file accessor for the local file manifest, figure out that the version on machine A is newer, and then compute which chunks of the file it should request from machine A.
The broadcast message from machine A included the process identifier of the specific file tracker on machine A, so the equivalent file tracker on machine B will now request the missing chunks directly from the file tracker on machine A.
This process identifier is only assumed valid for this one exchange, and is not retained on machine B for subsequent use.
Once the chunks have been delivered, the file tracker on machine B instructs its file accessor to rebuild the file according to the new file manifest.
It will supply the new chunks, and together with the chunks that already exist on machine B, the file is updated to have the same contents as the version on machine A.

\Cref{fig:sync-flow} shows two machines for clarity.
However, in general, a \ms{} instance may be connected to several remote peers.
The peer process broadcasts file-version announcements to all of them, and each receiving instance independently decides how to react to the announcement.
The described application is built entirely from \mch{} primitives such as named processes for discovery, message passing for routing, generic servers for stateful services, supervisors for recovery, etc.
The following paragraph gives some details on the implementation, but naturally omits most of the application code.

\paragraph{The Process Tree}

We have already touched upon several processes that are part of a \ms{} instance, but here we give a complete description of the different kinds of processes.
A \ms{} instance is implemented as a process tree, with the top-level process being the \ms{} process.
It registers itself under the name \inlinehaskell{"microsync"} and starts a supervisor for the long-running services of the application.
The supervisor uses a one-for-one strategy to permanently supervise four child processes.
The child processes in question are the \textit{Event Logger} process, the \textit{Folder Manager} process, the \textit{Peer Communication} process, and finally the \textit{Peer Connector} process.

The event logger is a generic server, registered under the name \inlinehaskell{"eventlogger"}, whose service is to accept events and log them to a persistent file on disk.
It gives the rest of the application a single named process to which events can be sent to for logging.

The peer connector process discovers remote \ms{} instances by querying connected \mch{} nodes for a registered \inlinehaskell{"peer"} process.
This peer process is the control-plane process for remote \ms{} instances: it tracks connected peers, monitors them for disconnection, and forwards file-version announcements between machines.

The folder manager process owns the state of the synchronized folder.
It manages the set of files being tracked at a point in time, and starts or stops file tracker processes as necessary.
Additionally, it manages a folder scanner process, which scans the folder for new and deleted files, while each tracked file is represented by a dynamically added file tracker process.

A file tracker process is created dynamically for each tracked file.
It owns the synchronization protocol for that file and presents the rest of the application with a single process to address.
The tracker itself is implemented as a generic server, registered under the file path, and it starts a private supervision tree for the helper processes associated with the file.
Its modify detector reports suspected local changes, while its file accessor serializes all reads and writes to the underlying file.

\begin{figure}
  \centering
  \includegraphics[scale=0.5]{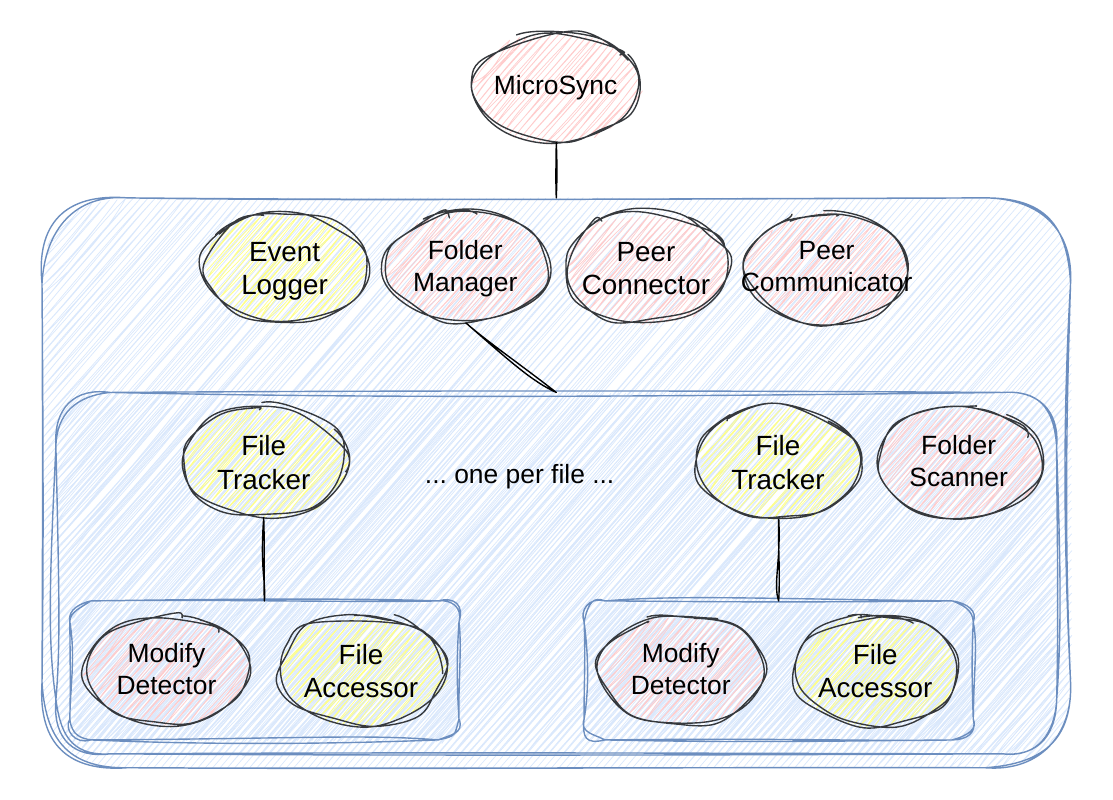}
  \caption{Process tree of one \ms{} instance. Blue containers are supervisors, yellow processes are generic servers, and red processes are ordinary processes. File trackers are created dynamically, one per tracked file, and each tracker supervises its own modify detector and file accessor.}
  \label{fig:processtree}
\end{figure}

\Cref{fig:processtree} illustrates how the processes are structured, and who supervises whom.

\subsection{Case Study 2: Heterogeneous Nodes}

The second case study exercises a different motivation for building \mch{} on MicroHs: portability rather than performance.
We run a \mch{} node on each of two STM32H563ZI microcontrollers running Zephyr RTOS, and a third node on the laptop used for the ring benchmark.
The microcontrollers have 2 MB of flash memory, 640 KB of SRAM, and run at up to 250 MHz.
All three devices are connected by Ethernet through a network switch.

This setup is deliberately heterogeneous.
The laptop node runs a 64-bit MicroHs runtime, whereas the microcontroller nodes run a 32-bit runtime built for Zephyr.
The nodes are therefore not executing the same binary, but they do share a compatible \mch{} protocol, graph format, and runtime representation.
This exercises the claim that \mch{} nodes need not agree on a common set of static symbols in the style of \ch{}.
Instead, they must agree on the runtime format used to serialize and deserialize graphs.
The messages used in this case study contain primitive values that are representable on both the 64-bit and 32-bit runtimes. The case study therefore demonstrates portability of the runtime format and programming model across heterogeneous nodes, not that every possible value of every architecture-dependent primitive can be transported unchanged.

The application is small but end-to-end.
One microcontroller runs a generic server that controls LEDs.
The second microcontroller runs a process that reacts to button presses and sends requests to the LED server.
The laptop runs a third \mch{} node that accepts keyboard commands and sends the same kind of requests.
Thus, both an embedded node and a laptop node communicate with a server process running on another embedded node, using the ordinary \mch{} message-passing interface.

This case study is not a performance benchmark.
Its purpose is to show that the implementation is not tied to GHC-sized machines or to a homogeneous cluster of identical binaries.
The same process abstraction, message format, and generic-server library used in the earlier examples can also be deployed across resource-constrained devices.

\section{Discussion}

\paragraph{Making Serialization Safer}
The implementation of \mch{} uses the unsafe primitives for serialization and deserialization.
This should not be read as the authors claiming that static checks are undesirable.
The goal of this project is to explore a programming model where the static checks by \ch{} are turned into runtime checks.

A safer API can be built on top of the unsafe primitives, eliminating some of the risk.
A natural refinement is to add a lightweight type-description constraint, for example:
\begin{lstlisting}[style=mchhaskell]
  serializeHL   :: TypeDescribable a => a -> IO ByteString
  deserializeHL :: TypeDescribable a => ByteString -> IO (Maybe a)
\end{lstlisting}
This constraint would not be used to define how to serialize a value.
That task is still delegated to the runtime system.
Instead, the constraint would be used to advertise the boundary type, so that a receiver can reject deserialization of a serialized value where the boundary types don't match.
Descriptors like this could be derived generically, for instance via \inlinehaskell{Data.Data}, and be made to include additional structural information.
This is a more explicit version of the check \mch{} already performs for messages: remote messages are sent along with a \inlinehaskell{TypeRep}, and selective receive compares that representation before deserializing the message payload.

This would sit between the solution of \ch{} and \mch{}.
There is no need for manual closure conversion or for the programmer to write serializers, but the communication boundary still becomes explicit in the source program.
This solution would not rule out serialization of runtime-owned resources such as \inlinehaskell{MVar}s or foreign pointers, but it would turn some type mismatches that are currently detected only at use sites into dynamic checks at deserialization time.
These safer functions would still enable the programmer to accidentally serialize thunks or unintentionally large graphs, however.

\paragraph{Benchmarks}

We begin our discussion by observing that the slow path takes significantly longer, increasing the required time from 60 \si{\micro\second} to some 550 \si{\micro\second}.
Of this, roughly 30 \si{\micro\second} is added by going through the node loop, and roughly 450 \si{\micro\second} comes from serialization and deserialization.
This is not unexpected, but ideally the cost would be smaller.
The serializer is implemented via two passes to preserve sharing and cycles.
A one-pass serializer would be faster, but would not be as capable.
We would not be able to serialize recursive graphs, and shared sub-graphs would be replicated as many times as they were referenced, increasing the size of the reconstructed graph.

The serialized process closure occupies roughly 32 KB.
The serialized graph is emitted in a textual ASCII format, which is not particularly compact.
A new output format could drastically reduce the size of the serialized graph.
There would perhaps be a slight gain in performance, but nothing extraordinary.

The results for the ring benchmark indicate that \mch{} is significantly slower than Erlang.
There are several reasons for this, but a notable one is that Erlang has been optimized for this very task over several decades, whereas \mch{} is running on MicroHs, a non-optimizing compiler in its infancy.
Additionally, even if MicroHs were an optimizing compiler, the underlying choice of reducing a mutable combinator graph at runtime will always suffer performance penalties.
While not reported in this paper, the authors point out that MicroHs's runtime is generally around two orders of magnitude slower than GHC's.

The heterogeneous-node case study explains why this tradeoff may still be attractive.
MicroHs is much slower than mature native-code systems, but its runtime is small and portable.
In our prototype, the compiled runtime is roughly 300 KB, small enough to run on the STM32H563ZI microcontrollers used in the case study.
The case study therefore supports a portability claim rather than a performance claim: \mch{} makes it possible to use the same distributed Haskell programming model across machines that would not normally be able to run a GHC-based distributed Haskell system.

The distributed prime counter shows significant speedups compared to sequential runtime.
We point out that there seems to be a superlinear speedup when going from one to two remote machines, and we are not entirely sure why.
The remote instances are separate instances, but it is impossible to say precisely where they are hosted and what else is hosted on the same hardware.
While we cannot say for sure, we speculate that the phenomenon may be due to caching, and that using more than one MicroHs instance reduces the frequency with which the garbage collector has to run.

\paragraph{\ms{}} 

The purpose of \ms{} is not to evaluate file synchronization performance, but to test whether the \mch{} programming model is usable in a larger application.
The case study exercises many things, including named processes, dynamic process creation, supervisors, monitors, generic servers, and cross-node message passing in one program.
It therefore supports an expressiveness claim rather than a performance claim, namely that \mch{} is sufficient to structure a nontrivial distributed application while preserving the direct style illustrated in the smaller examples.

\section{Related Work}

\paragraph{\ch{}}
The work described in this \paper{} is strongly related to \ch{}.
\ch{} presented an elegant API, as evidenced by the work described in this \paper{} reusing many parts of it.
The primary difference is the treatment of \inlinehaskell{spawn}, where \ch{} required the programmer to manually apply lambda-lifting to their process bodies.
Furthermore, the programmer had to apply explicit closure conversion when spawning such functions, adding cognitive overhead.
\mch{} removes this source-level closure-conversion step.
A process body can be written directly at the call to \inlinehaskell{spawn}, including references to variables in scope, and the runtime serializes the reachable graph when the process crosses a node boundary.

\ch{} explicitly discusses the style of implementation used by \mch{}, where the compiler or runtime provides primitives for serializing arbitrary values, and argues against it.
The concerns raised there still apply.
Some runtime-owned values, such as open file handles, should not be serializable.
A programmer may also accidentally serialize a much larger graph than intended, or serialize an unevaluated computation whose work was expected to happen before transmission.

\paragraph{Runtime-Level Code Mobility}
\mch{} relies on a runtime representation in which code and data can both be serialized as graph nodes.
This approach is less natural for implementations such as GHC, which compile Haskell to native machine code.
Machine code is neither portable nor desirable as a communication format.
The same idea is more plausible for implementations based on bytecode or another portable intermediate representation.
In such systems, the code representation can be included in the serialized data and reconstructed, interpreted, or JIT-compiled by the receiver.
This is how the Mu implementation at Standard Chartered worked~\cite{Augustsson2011}.

\paragraph{Glasgow Distributed Haskell}
Glasgow Distributed Haskell~\cite{DBLP:conf/ifl/PointonTL00} (GDH) was an earlier effort to extend Haskell for distributed programming.
GDH preceded \ch{}, but the general idea was the same, that a distributed application should be described as a collection of communicating processes.
GDH was elegant and well-designed, but practical concerns led to the project slowing down.
It required substantial modification to GHC's runtime system of the time (the GUM~\cite{trinder1996gum} runtime), which limited its adoption and maintainability.

\paragraph{Elixir}
Elixir~\cite{elixir} runs on Erlang's runtime, retaining its actor model while offering more approachable syntax, modern tooling, and a modern ecosystem.

\paragraph{Tierless and Choreographic Haskell}
Recent work has explored several ways to describe distributed Haskell programs from a single source program.
In tierless programming, the behaviors of multiple tiers are written and type-checked together, then compiled into separate executables.
The framework of Ekblad and Claessen~\cite{DBLP:conf/haskell/EkbladC14} follows this style for client-server web applications.
The client and server are described in one program, and the compiler separates the code so that client code does not appear in the server, and vice versa.

The idea was later adopted by HasTEE~\cite{DBLP:conf/haskell/SarkarKRC23}, a framework for programming \textit{Trusted Execution Environments} in Haskell.
They observed that a trusted computing unit running on Intel SGX hardware assumes the role of a server, and that the programming model of Ekblad and Claessen~\cite{DBLP:conf/haskell/EkbladC14} was a good fit because of its separation of client code and server code.

Choreographic programming~\cite{montesi2014choreographic} takes a related whole-program view, but describes communication protocols globally rather than from one endpoint's perspective.
A communication action is written once, naming both sender and receiver.
When the choreography is interpreted or projected for a particular endpoint, that action becomes a send, a receive, or no local action.
HasChor~\cite{DBLP:journals/pacmpl/ShenKK23} embeds this style of choreographic programming in Haskell.
Its presentation is elegant and simple, but the original implementation produced executables that contained the code for every endpoint in the network.
Krook and Hammersberg~\cite{DBLP:conf/haskell/KrookH24} address this limitation by specializing HasChor programs with respect to a known endpoint, producing executables that contain only the code needed by that endpoint.

\section{Conclusions \& Future Work}

We set out to explore what the \ch{} programming model looks like when serialization is provided by the runtime system rather than exposed as \inlinehaskell{Closure} and \inlinehaskell{Serializable} constraints in the source-level API.
\mch{} shows that this design point is practical: remote processes can be spawned in direct style, messages can contain ordinary values and functions, and familiar Erlang-style abstractions such as generic servers and supervisors can be implemented as ordinary libraries.

The benefit is ergonomic: programs can be written much like local communicating processes, with the runtime serializing the reachable MicroHs graph when computation or data crosses a node boundary.
The cost is that this boundary moves to runtime: deserialization depends on the receiver's expected type, and programmers must avoid moving runtime-owned resources or unevaluated work unintentionally.

Future work should reduce what crosses the wire.
One direction is to give nodes a shared vocabulary of known code and send references where possible, falling back to graph serialization for dynamic code and environments.
Another is to move low-level spawning and message delivery into the MicroHs runtime, reducing overhead while centralizing dynamic checks for graph types, runtime-owned resources, and version compatibility.




\begin{acks}
  The authors would like to thank Koen Claessen, Mary Sheeran, John Hughes, and Simon Peyton-Jones for their helpful discussions and feedback.
  We also thank Jessica Augustsson for editing and the anonymous reviewers for their helpful suggestions.
\end{acks}

\newpage
\appendix

\section{API Reference}

\label{appendix:API}

This appendix lists the core \mch{} API used throughout the \paper{}.

\begin{lstlisting}[style=mchhaskell]
  data Node a
  data NodeId
  data NodeConfig -- = (ip, port)
  data ProcessM a; instance MonadIO ProcessM
  data Pid
  runNode :: NodeConfig -> Node () -> IO ()
\end{lstlisting}

A node id says \textit{where} something is: a location in a network.
When a node connects to an existing network of nodes, they form a mesh, whereby every node is connected to every other node.

\begin{lstlisting}[style=mchhaskell]
  connect   :: NodeConfig -> ProcessM ()
\end{lstlisting}

Each node can host zero or more processes, which are independent computation units.
Processes (almost) don't share any memory with other processes, with the exception(s) described in \cref{sec:implementation}.
There are some basic operations for processes to identify themselves, their node, the nodes that are connected to the network, and to terminate.

\begin{lstlisting}[style=mchhaskell]
  self      :: ProcessM Pid
  node      :: ProcessM NodeId
  nodes     :: ProcessM [NodeId]
  terminate :: ProcessM a
\end{lstlisting}

A process that wishes to retrieve a list of every node in the network, aside from its own host node, may execute

\begin{lstlisting}[style=mchhaskell]
  do mynode <- node
     allnodes <- nodes
     return $ filter (/= mynode) allnodes
\end{lstlisting}

Processes are started either by running an initial process on a node, or by spawning a new process from an existing one.
A derived \inlinehaskell{call} operation spawns a process and waits for its result.

\begin{lstlisting}[style=mchhaskell]
  runProcessM :: ProcessM () -> Node ()
  spawn :: NodeId -> ProcessM () -> ProcessM Pid
  call  :: NodeId -> ProcessM a -> ProcessM a
\end{lstlisting}

As the processes are independent and don't share any memory, they cannot perform any meaningful task without interacting with each other.
Interactions are done via message passing, where processes can \inlinehaskell{send} and \inlinehaskell{expect} messages to and from one another.
\inlinehaskell{send} is an asynchronous call that returns immediately, whereas \inlinehaskell{expect} is a synchronous call that doesn't return until it has received a message of the correct type.
It is appropriate to think of it as if every process had a mailbox where messages of any type can go in, and where \inlinehaskell{expect} goes through the letters in the order they arrived, looking for one of the right type.

\begin{lstlisting}[style=mchhaskell]
  send   :: Identifiable process => process -> a -> ProcessM ()
  expect :: ProcessM a
\end{lstlisting}

For selective receive, \mch{} follows the matcher interface introduced by \ch{}~\cite{DBLP:conf/haskell/EpsteinBJ11}.
A matcher specifies one acceptable message type, optionally guarded by a predicate, and \inlinehaskell{receiveWait} blocks until any matcher succeeds.
There is additionally a \inlinehaskell{matchUnknown} handler that will match a message of any type, removing it from the mailbox.

\begin{lstlisting}[style=mchhaskell]
  data MatchM q a
  match          :: (a -> ProcessM q) -> MatchM q ()
  matchIf        :: (a -> Bool) -> (a -> ProcessM a) -> MatchM q ()
  matchUnknown   :: ProcessM q -> MatchM q ()
  receiveWait    :: [MatchM q ()] -> ProcessM q
  receiveTimeout :: Int -> [MatchM q ()] -> ProcessM (Maybe q)
\end{lstlisting}

A matcher is one branch of a selective receive.
It recognizes a message and runs a handler producing a value of type \inlinehaskell{q}.
\inlinehaskell{receiveWait} blocks until one of the supplied matchers succeeds, while \inlinehaskell{receiveTimeout} bounds the wait and returns \inlinehaskell{Nothing} if no message matches in time.

For example, a process can wait for either a request to divide two numbers, or for a command to shut down

\begin{lstlisting}[style=mchhaskell]
  data Div       = Div Pid Int Int
  data Shutdown = Shutdown

  server :: ProcessM ()
  server = receiveWait
    [ matchIf (\(Div _ _ y) -> y /= 0) $ \(Div sender x y) ->
        send sender (x `div` y)
    , match $ \Shutdown -> terminate
    ]
\end{lstlisting}

The two matchers receive messages of different types, but both branches are part of the same blocking receive.
The first branch accepts only non-zero division requests and replies to the sender.
The second branch accepts a shutdown command and terminates the process.
Messages that do not match either branch remain in the mailbox.

So far, the API describes how live processes communicate.
For fault-tolerant programs, processes must also be able to observe one another's lifetime, where a missing reply may mean that another process has terminated, crashed, or disappeared with its node.
\mch{} exposes this through monitors.

\begin{lstlisting}[style=mchhaskell]
  data ProcessDied   = ProcessDied Pid ExitReason
  data MonitorAction = TrapExit | Succumb

  monitor :: MonitorAction -> Pid -> ProcessM ()
\end{lstlisting}

When process \inlinehaskell{p} monitors process \inlinehaskell{q}, \inlinehaskell{p} is notified if \inlinehaskell{q} terminates.
With \inlinehaskell{TrapExit}, the notification is delivered as a \inlinehaskell{ProcessDied} message in \inlinehaskell{p}'s mailbox.
With \inlinehaskell{Succumb}, \inlinehaskell{p} terminates as well.

Aside from observing process termination, processes can also cause process termination, using the \inlinehaskell{exit} function.
This makes it possible to designate application-specific roles to processes, such as workers and supervisors.
Supervisor processes might send an exit signal to a process that has entered an unhealthy state.

\begin{lstlisting}[style=mchhaskell]
  data ExitReason
    = ExitNormal       -- ^ Process exited normally
    | ExitShutdown     -- ^ Process exited upon request of another
                       --   process, e.g. a supervisor
    | ExitKill         -- ^ Process was forcibly killed, equivalent to kill -9.
    | ExitOther String -- ^ Other custom reasons why a process was terminated

  exit :: Pid -> ExitReason -> ProcessM ()
\end{lstlisting}

An example that illustrates how \inlinehaskell{monitor} and \inlinehaskell{exit} can be used to control behavior is shown below

\begin{lstlisting}[style=mchhaskell]
  do  child <- spawn n childBody
  
      _ <- spawn n $ do
        monitor TrapExit child
        ProcessDied _ _ <- expect
        liftIO (putStrLn "the child died!")
  
      linked <- spawn n $ do
        monitor Succumb child
        expect
  
      monitor TrapExit linked
      exit child ExitKill
      ProcessDied _ _ <- expect
\end{lstlisting}
  
The first monitor survives the child's death and receives a \inlinehaskell{ProcessDied} message.
The second monitor uses \inlinehaskell{Succumb}, so it dies together with the child.
By monitoring the second process with \inlinehaskell{TrapExit}, the parent observes this cascading termination as an ordinary mailbox message.

The process registry maps node-local names to process identifiers.
In the example above, the two monitoring processes had to have a way of knowing the process identifier of the child process.
Sometimes it is not ergonomic to distribute process identifiers to everyone who has to know them.
Furthermore, a process may be restarted by a supervisor process, in which case the pre-distributed process identifier will no longer be valid.
The process registry lets a process register itself under a name on the node it is running on.

\begin{lstlisting}[style=mchhaskell]
  register   :: Pid -> String -> ProcessM ()
  unregister :: String -> ProcessM ()
  whois      :: String -> ProcessM (Maybe Pid)
\end{lstlisting}

Since \inlinehaskell{String} is an instance of \inlinehaskell{Identifiable}, \inlinehaskell{send} may target a registered name.

\begin{lstlisting}[style=mchhaskell]
  send :: Identifiable i => i -> a -> ProcessM ()
\end{lstlisting}

The \inlinehaskell{Identifiable} instances used here are \inlinehaskell{Pid} and \inlinehaskell{String}.
If a message is sent to a named process, the process identifier is fetched from the process registry before the message is sent.
If the name is not registered, the \inlinehaskell{send} fails with an exception, much like it would in Erlang.

We emphasize that the name registry is node-local, meaning that messages can only be sent to named processes on the same node.
The process identifier of a remote named process can be retrieved without much fuss, however.
We can simply issue a synchronous call, reading the remote registry

\begin{lstlisting}[style=mchhaskell]
  remoteProcess <- call n' (whois "child") -- n' is the NodeId of some remote node
  case remoteProcess of
    Just pid -> send pid "hello"
    Nothing -> return ()
\end{lstlisting}

\newpage
\bibliographystyle{ACM-Reference-Format}
\bibliography{main}

\end{document}